%% file: 0_Main.tex
\documentclass[sigconf]{acmart}
\input{macros}

\usepackage[export]{adjustbox}
\usepackage{subcaption}
\usepackage{enumitem}
\usepackage{makecell}
\usepackage{array,multirow}

\newsavebox{\measurebox}

\newcommand{\gctr}{\text{gCTR}}
\newcommand{\pctr}{\text{pCTR}}
\newcommand{\ghr}{\text{gHR}}
\newcommand{\phr}{\text{pHR}}
\newcommand{\refClick}{\text{\textit{refClick}}}
\newcommand{\extClick}{\text{\textit{extClick}}}
\newcommand{\fnHover}{\text{\textit{fnHover}}}
\newcommand{\fnClick}{\text{\textit{fnClick}}}
\newcommand{\upClick}{\text{\textit{upClick}}}
\newcommand{\pageLoad}{\text{\textit{pageLoad}}}
\newcommand{\END}{\text{\textit{END}}}

\definecolor{airforceblue}{rgb}{0.2, 0.2, 0.66}

\copyrightyear{2020}
\acmYear{2020}
\setcopyright{iw3c2w3}
\acmConference[WWW '20]{Proceedings of The Web Conference 2020}{April 20--24, 2020}{Taipei, Taiwan}
\acmBooktitle{Proceedings of The Web Conference 2020 (WWW '20), April 20--24, 2020, Taipei, Taiwan}
\acmPrice{}
\acmDOI{10.1145/3366423.3380300}
\acmISBN{978-1-4503-7023-3/20/04}

\title{Quantifying Engagement with Citations on Wikipedia}

\begin{document}

\author{
\authorbox{Tiziano Piccardi}{EPFL}{tiziano.piccardi@epf\/l.ch}\hspace{7mm}
\authorbox{Miriam Redi}{Wikimedia Foundation}{miriam@wikimedia.org}\hspace{7mm}
\authorbox{Giovanni Colavizza}{University of Amsterdam}{g.colavizza@uva.nl}\hspace{7mm}
\authorbox{Robert West*}{EPFL}{robert.west@epf\/l.ch}
}

\renewcommand{\shortauthors}{Tiziano Piccardi, Miriam Redi, Giovanni Colavizza, and Robert West}


\begin{abstract}
Wikipedia is one of the most visited sites on the Web and a common source of information for many users. As an encyclopedia, Wikipedia was not conceived as a source of original information, but as a gateway to secondary sources: according to Wikipedia's guidelines, facts must be backed up by reliable sources that reflect the full spectrum of views on the topic. Although citations lie at the heart of Wikipedia, little is known about how users interact with them. To close this gap, we built client-side instrumentation for logging all interactions with links leading from English Wikipedia articles to cited references during one month, and conducted the first analysis of readers' interactions with citations. We find that overall engagement with citations is low: about one in 300 page views results in a reference click (0.29\% overall; 0.56\% on desktop; 0.13\% on mobile). Matched observational studies of the factors associated with reference clicking reveal that clicks occur more frequently on shorter pages and on pages of lower quality, suggesting that references are consulted more commonly when Wikipedia itself does not contain the information sought by the user. Moreover, we observe that recent content, open access sources, and references about life events (births, deaths, marriages, \etc)\ are particularly popular. Taken together, our findings deepen our understanding of Wikipedia's role in a global information economy where reliability is ever less certain, and source attribution ever more vital.
\end{abstract}

\maketitle

{\fontsize{8pt}{8pt} \selectfont
\textbf{ACM Reference Format:}\\
Tiziano Piccardi, Miriam Redi, Giovanni Colavizza, and Robert West.
2020.
Quantifying Engagement with Citations on Wikipedia.
In
\textit{Proceedings of The Web Conference 2020 (WWW '20), April 20--24, 2020, Taipei, Taiwan.}
ACM, New York, NY, USA, 12 pages. \url{https://doi.org/10.1145/3366423.3380300}
}

\blfootnote{*Robert West is a Wikimedia Foundation Research Fellow.}

\input{1_Introduction}
\input{2_Related}
\input{3_Methods}
\input{4_RQ1}
\input{5_RQ2}

\input{6_RQ3}
\input{7_Discussion}

\xhdr{Acknowledgments}
We thank Leila Zia, Michele Catasta, Dario Taraborelli for early contributions; Bahodir Mansurov, WMF Analytics for help with event logging;
James Evans for good discussions;
Microsoft, Google, Facebook, SNSF for supporting West's lab.

\balance

\bibliographystyle{ACM-Reference-Format}
\bibliography{citationusage}

\end{document}

%% file: macros.tex
\usepackage{hyphenat}
\usepackage{xspace}
\usepackage{amsmath} 
\usepackage{multirow}
\usepackage{makecell}
\usepackage{caption}
\usepackage{minibox}
\usepackage{bbm}
\usepackage{graphicx}
\usepackage{balance}
\usepackage[show]{chato-notes} 

\newcommand{\textcite}[1]{\citeauthor{#1} \shortcite{#1}}

\newcommand{\ie}{{i.e.}\xspace}
\newcommand{\eg}{{e.g.}\xspace}
\newcommand{\cf}{{cf.}\xspace}

\newcommand{\vs}{{vs.}\xspace}
\newcommand{\etc}{{etc.}\xspace}

\newcommand{\Secref}[1]{Sec.~\ref{#1}}
\newcommand{\Eqnref}[1]{Eq.~\ref{#1}}

\newcommand{\Tabref}[1]{Table~\ref{#1}}
\newcommand{\Figref}[1]{Fig.~\ref{#1}}

\newcommand{\hide}[1]{}

\newcommand{\xhdr}[1]{\vspace{1.7mm}\noindent{{\bf #1.}}}

\newcommand{\affilSize}{9pt}
\newcommand{\authorbox}[3]{
\minibox[c]{
#1\\
{\fontsize{\affilSize}{\affilSize}\selectfont{}#2}\\
{\fontsize{\affilSize}{\affilSize}\selectfont{}#3}
}
}

\newcommand\blfootnote[1]{%
  \begingroup
  \renewcommand\thefootnote{}\footnote{#1}%
  \addtocounter{footnote}{-1}%
  \endgroup
}

%% file: 1_Introduction.tex
\section{Introduction}
Wikipedia is the largest encyclopedia ever built,
established through the collaborative effort of a large editor base, self-governed through agreed policies and guidelines \cite{beschastnikh_wikipedian_2008,forte_decentralization_2009}. Thanks to the tenacious work of the editor community, Wikipedia's content is generally up to date and of high quality \cite{keegan_hot_2011,piscopo_what_2019}, and is relied upon as a source of neutral, unbiased information \cite{mesgari_sum_2015}.

\begin{figure}
    \centering
    \includegraphics[width=\columnwidth]{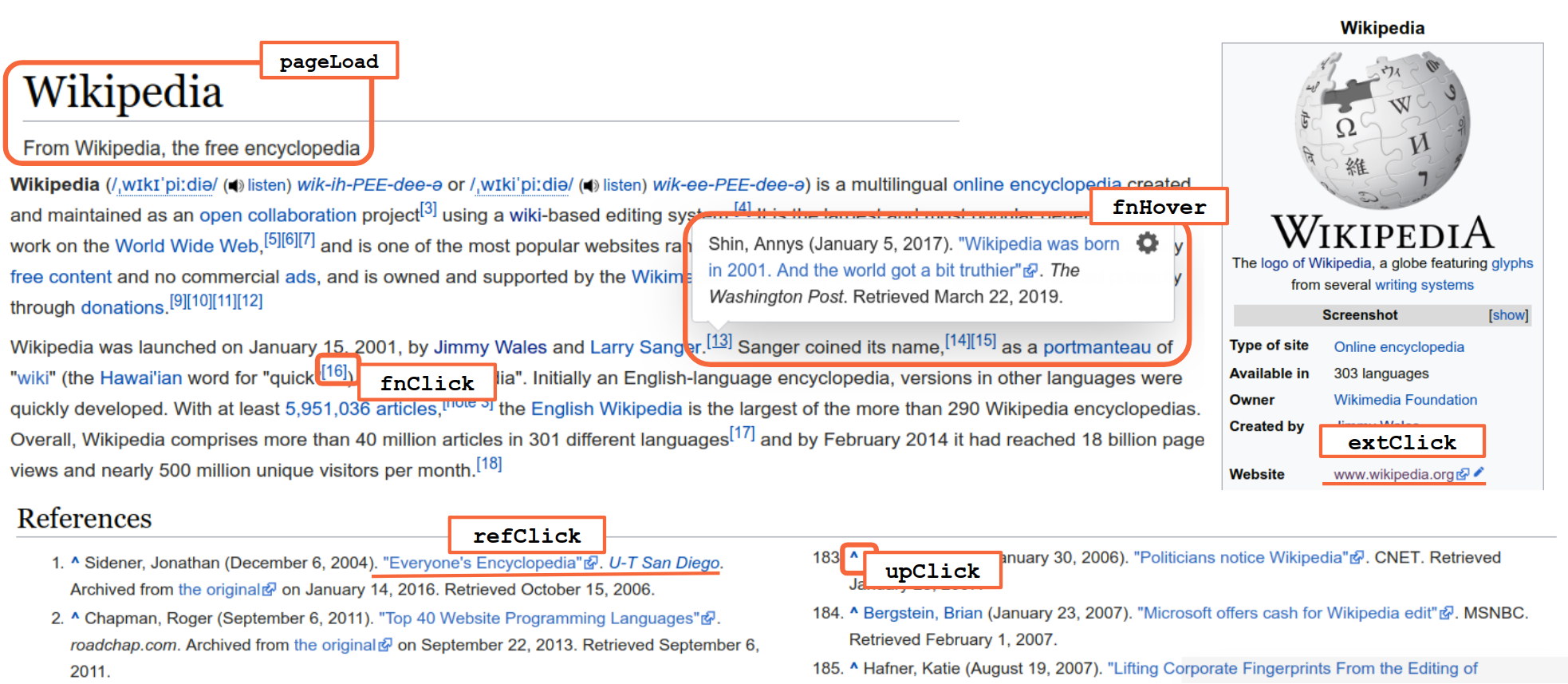}
    \caption{Examples of the 6 types of interactions with pages and citations that we record on English Wikipedia using Wikimedia's EventLogging tool.}
    \label{fig:event_types}
\end{figure}

Wikipedia's inline references, or citations,%
\footnote{
We use the terms ``reference'' and ``citation'' largely interchangeably.
}
are a key mechanism for monitoring and maintaining its high quality. Wikipedia's core content policies require that ``people using the encyclopedia can check that the information comes from a reliable source'',%
\footnote{
\url{https://en.wikipedia.org/wiki/Wikipedia:Verifiability}, \url{https://en.wikipedia.org/wiki/Wikipedia:Reliable_sources}
}
and citations are the main way to connect a statement to its sources. A clearly distinctive feature of Wikipedia is the fact that many citations are actionable: they are often equipped with hyperlinks to the cited material available on the Web.

As a result, Wikipedia's role on the Web has been defined as the ``bridge to the next layer of academic resources'' \cite{grathwohl2011wikipedia}, and the ``gateway through which millions of people now seek access to knowledge'' \cite{gateway}.
Nevertheless, a question remains open: to which extent do Wikipedia readers actually cross the bridge and access the broader knowledge referenced in the encyclopedia? 

Given the collaborative and open nature of Wikipedia, being able to quantify readers' engagement with the content and its supporting sources is of crucial importance for the constant betterment of the encyclopedia and its role in fostering a self-critical society. By understanding readers' interactions with citations, we can better assess the role of Wikipedia editors and policies in maintaining a high quality of information,
measure public demand for secondary sources,
and provide insights and potential recommendations to increase the public's interest in references.

This paper takes a step in this direction, by addressing, for the first time, the problem of quantifying and studying Wikipedia readers' engagement with citations.
More specifically, we  ask the following research questions,
\begin{description}
    \item[RQ1] To what extent do users engage with citations when reading Wikipedia? (\Secref{sec:RQ1})
    \item[RQ2] What features of a page predict whether a reader will interact with a citation on the page? (\Secref{sec:RQ2})
    \item[RQ3] What features of a citation predict whether a reader will interact with it? (\Secref{sec:RQ3})
\end{description}

In order to answer these questions, we collect a large dataset comprising all citation\hyp related events (96M) on the English Wikipedia for two months (October 2018, April 2019),
including reference clicks, reference hovers, and downwards and upwards footnote click, as visualized in \Figref{fig:event_types}.
By analyzing this dataset,%
\footnote{Notebooks with code at \url{https://github.com/epfl-dlab/wikipedia-citation-engagement}}
we make the following main contributions:
\begin{itemize}[leftmargin=*]
    \item We quantify users' engagement with citations and find that it is a relatively rare event (RQ1, \Secref{sec:RQ1}): 93\% of the links in citations are never clicked over a one-month period, and the fraction of page views that involve a click on a citation link is $0.29\%$.
    \item We gain insights into factors associated with seeking additional information via citation interactions, both at the page level (RQ2, \Secref{sec:RQ2}) and at the link level (RQ3, \Secref{sec:RQ3}). Through matched observational studies, we show that articles that are of higher quality, and thus also longer and more popular, are associated with a lower propensity of users to interact with citations. Using a logistic regression model trained on linguistic features, we show that more frequently clicked citation links tend to relate to social or life events.
\end{itemize}

We thus conclude that readers are more likely to use Wikipedia as a \textit{gateway} on topics where Wikipedia is still wanting and where articles are of low quality and not sufficiently informative; and that Wikipedia tends to be the \textit{final destination} in the large majority of cases where the information it contains is of sufficiently high quality.
 
Our work provides the first study aimed at understanding if and how users engage with citations on Wikipedia, thus paving the way for a broader and deeper understanding of Wikipedia's role in the global information ecosystem.

%% file: 2_Related.tex
\section{Related Work}
This paper is related to research on a number of different themes.

\xhdr{Characterizing Wikipedia readers}
A substantial amount of prior work has focused on understanding user engagement with Wikipedia from the point of view of the \textit{editor} community \cite{oded_what_2007,yasseri_dynamics_2012,arazy_how_2017,petrasova_similar_2018}. Studies on the behavior of Wikipedia \textit{readers} have mostly considered interest in contents \cite{waller_search_2011,lehmann_reader_2014,salutari_large-scale_2019}, content popularity \cite{spoerri_what_2007,ratkiewicz_characterizing_2010,chelsy_xie_detecting_2019}, or event timing \cite{moat_quantifying_2013}. More recently, a study explored the question why users read Wikipedia, by combining multiple-choice surveys with log-based analyses of user activity \cite{singer_why_2017}. A similar design was used to study 14 languages other than English \cite{lemmerich_why_2019}. Little is known, however, on how users engage with Wikipedia's citations of external sources; ours is the first study on this subject.

\xhdr{Navigation in Wikipedia}
Wikipedia citations are part of the hyperlink network connecting Wikipedia and the Web. Understanding citation usage can yield useful insights for improving this network \cite{west_mining_2015,lamprecht_evaluating_2016}. The analysis, modeling, and prediction of human navigation \textit{inside} Wikipedia has been considered in previous studies \cite{trattner_exploring_2012,helic_models_2013,singer_detecting_2014,lamprecht_how_2017,cornelius_inspiration_2018,dimitrov_different_2019}, largely relying on traces from the navigation games Wikispeedia \cite{west_human_2012,west_automatic_2012,scaria_last_2014} and WikiGame \cite{singer_computing_2013,dallmann_extracting_2016,koopmann_right_2019}. For our study, we collect instead a new, fine-grained dataset of user interactions with Wikipedia references to \textit{external} content.

\xhdr{Science in Wikipedia}
A sizeable portion of citations on Wikipedia refer to scientific literature \cite{blomqvist_scholia_2017}. Consequently, Wikipedia is a fundamental gateway to scientific results and enables the public understanding of science \cite{nielsen_scientific_2007,damasevicius_analysis_2017,shafee_evolution_2017,maggio_wikipedia_2019,torres-salinas_mapping_2019}.
The chance of a scientific reference being cited on Wikipedia varies with the impact factor of the publication venue and its open\hyp access availability \cite{teplitskiy_amplifying_2017}.
Being cited on Wikipedia can thus be considered an indicator of impact \cite{kousha_are_2017}.
Despite the indirect influence that Wikipedia has on scientific progress \cite{thompson_science_2018}, Wikipedia is in turn rarely acknowledged in the scientific literature \cite{jemielniak_bridging_2016,tomaszewski_study_2016}.

\xhdr{Improving Wikipedia}
Wikipedia content quality relies on the work of editors and their gradual improvement of articles \cite{priedhorsky_creating_2007,chen_citation_2012}.
Automated or semi\hyp automated tools \cite{nielsen_wikipedia_2012,geiger_when_2013,Piccardi:2018} can help improve user experience \cite{lamprecht_evaluating_2016,yazdanian_eliciting_2019}, content variety \cite{paranjape_improving_2016,wulczyn_growing_2016}, and quality \cite{adler_assigning_2008,kumar_disinformation_2016,halfaker_interpolating_2017}. The reliability of Wikipedia can also be improved automatically, \eg, by finding potential citations \cite{fetahu_finding_2016} and Wikipedia statements in need of evidence \cite{redi_citation_2019}. The insights from our work can help improve Wikipedia via new citations with which users would be more likely to interact.

\xhdr{Quantifying Web user engagement}
User engagement is crucial for the success of Web services, and numerous researchers have focused on quantifying how Web users engage with online content,
\eg, in computational advertising \cite{barbieri2016improving,yi2015dwell},
social media \cite{hu2015predicting,claussen2013effects,bakhshi2014faces}, or
information retrieval \cite{song2013evaluating,jing2015visual}.
Also, while the body of work focusing on understanding readers' and editors' engagement with content \textit{within} Wikipedia has been growing in the recent years \cite{miquel2015user}, we study here for the first time how Wikipedia readers engage with the broader \textit{outside} knowledge linked from the online encyclopedia.

%% file: 3_Methods.tex
\section{Citation data collection}\label{sec:Citation data collection}

To study readers' engagement with citations, we collected data capturing where readers navigate and how they interact with citations in English Wikipedia.

\subsection{Background: Citations in Wikipedia}
\label{sec:Background: Citations in Wikipedia}

Articles in Wikipedia are written by editors in wikicode, a markup language that is then translated to HTML by MediaWiki, the software that powers the website. There are different ways to add citations to sources in the text, summarized below. In all cases, the full reference descriptions are rendered as footnotes at the bottom of the page (in a dedicated section called \textit{References}) with an automatically assigned footnote number that is added as a link anchor (e.g., ``[1]'') in the text of the article wherever the reference is cited (\Figref{fig:event_types}). 
Most references in the \textit{References} section consist of text including the title of the source, the authors' names, the year of publication, and the source's publisher. For 80\% of Wikipedia references, the source title is actionable via a clickable link to the source.
Also, when reading a page, hovering over a reference's footnote number with the mouse cursor will display a \textit{reference tooltip},%
\footnote{\url{https://www.mediawiki.org/wiki/Reference_Tooltips}}
a pop-up containing the reference text and a clickable link (when present), \eg,
\begin{quote}
    Daniel Nasaw (July 24, 2012).  \href{https://www.bbc.co.uk/news/magazine-18892510}{\textcolor{blue}{``Meet the `bots' that edit Wikipedia''}}. BBC News.
\end{quote}
When readers click on the reference's footnote number, they are sent to the reference description at the page bottom, from where they can jump back to the locations where the reference is cited by clicking on a small icon (e.g., \verb!^!).

The most common method to add a reference to an article, also recommended by the Wikipedia guidelines, is via an inline citation using a \texttt{<ref/>} tag directly in the context where the reference is first cited. In the tag, the editors can specify the reference details (text and links) by using a predefined template or plain wikicode. In addition to this standard method, some references are added automatically by templates included in the page, such us the geolocations present in the infobox. It is worth noting that a reference can be cited multiple times by assigning it a name and appending the tag to every sentence that should link to it. Given the numerous ways to use the \texttt{<ref/>} tag, and in order to have an accurate view of the article, we parsed pages from wikicode to HTML and extracted the information from the HTML code.

\subsection{Logging citation and page load events}
\label{sec:Logging citation and page load events}

We make use of Wikimedia's \textit{EventLogging} tool,%
\footnote{\url{https://www.mediawiki.org/wiki/Extension:EventLogging/Guide}}
an extension of the MediaWiki software
that performs client-side logging of specific types of events.
We detect 5 main types of citation-related events and 1 page load event. In terms of citations, we capture the mouse events that involve any kind of reader interaction with the references (see \Figref{fig:event_types} for a visual explanation):
\begin{description}
\item[\textit{refClick:}] a click on a hyperlink in an article's reference section.
\item[\textit{extClick:}] a click on an external link outside the reference section.
\item[\textit{fnHover:}] a hover over a footnote number in the text, logged when the reference tooltip is visible for more than 1 second.
\item[\textit{fnClick:}] a click on a footnote number, which takes the user to the reference section at the bottom of the page.
\item[\textit{upClick:}] the inverse of \textit{fnClick:} a click on a reference's up arrow icon that takes the reader back to the part of text where the reference is cited.
\item[\textit{pageLoad:}] in addition to the above citation\hyp related events, this event is triggered whenever a Wikipedia article is loaded.
\end{description}

The EventLogging platform manages a so-called \textit{session token,} a cookie-based identifier that allows us to group events that happened within the same browser tab. We henceforth refer to event sequences that occur with the same session token as \textbf{sessions.}

We collected 4 contiguous weeks\footnote{We collected exactly 4 weeks to reduce potential seasonal effects due to uneven day-of-the-week frequencies.} of Wikipedia mobile and desktop traffic data of citation-related events. We repeated the 4-week data collection over two periods: from September 26 to October 25, 2018, and from March 24 to April 21, 2019. In both cases, we collected all citation-related events (\extClick, \refClick, \fnHover, \fnClick, \upClick)
and (due to computational infrastructure constraints) sampled \pageLoad{} events at the session level at a rate of 33\%.

To ensure that the logs reflect reader, rather than editor, behavior, we exclusively retained data from users who in the 4 weeks of data collection acted only as anonymous readers, discarding all events generated by Wikipedia editors (logged-in users or users with anonymous edits) and by bots (which can be filtered out using a detector provided by the EventLogging tool).

\begin{figure*}[t]
 \begin{subfigure}[b]{0.33\textwidth}
        \includegraphics[width=\textwidth,left]{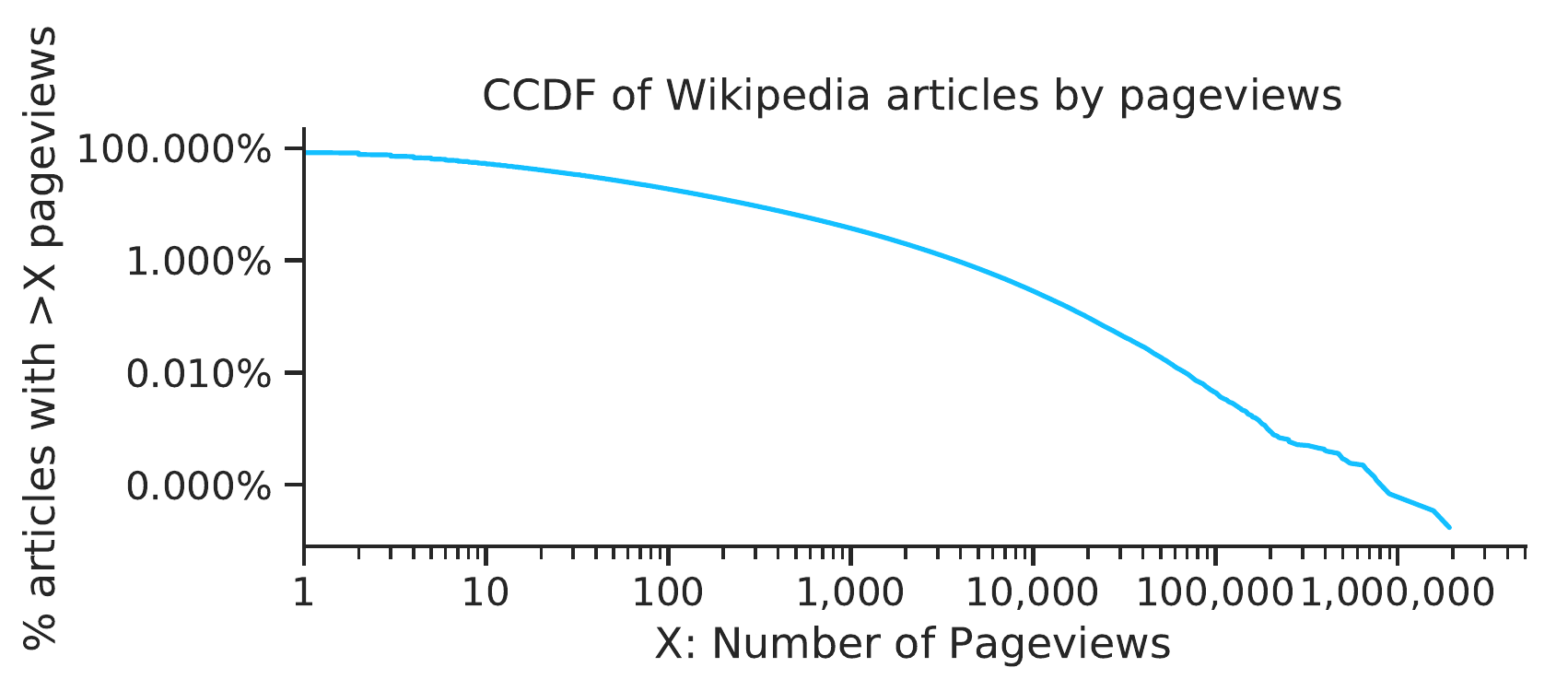}
        \caption{}
        \label{fig:pageviews}
    \end{subfigure}
     \begin{subfigure}[b]{0.33\textwidth}
        \includegraphics[width=\textwidth,right]{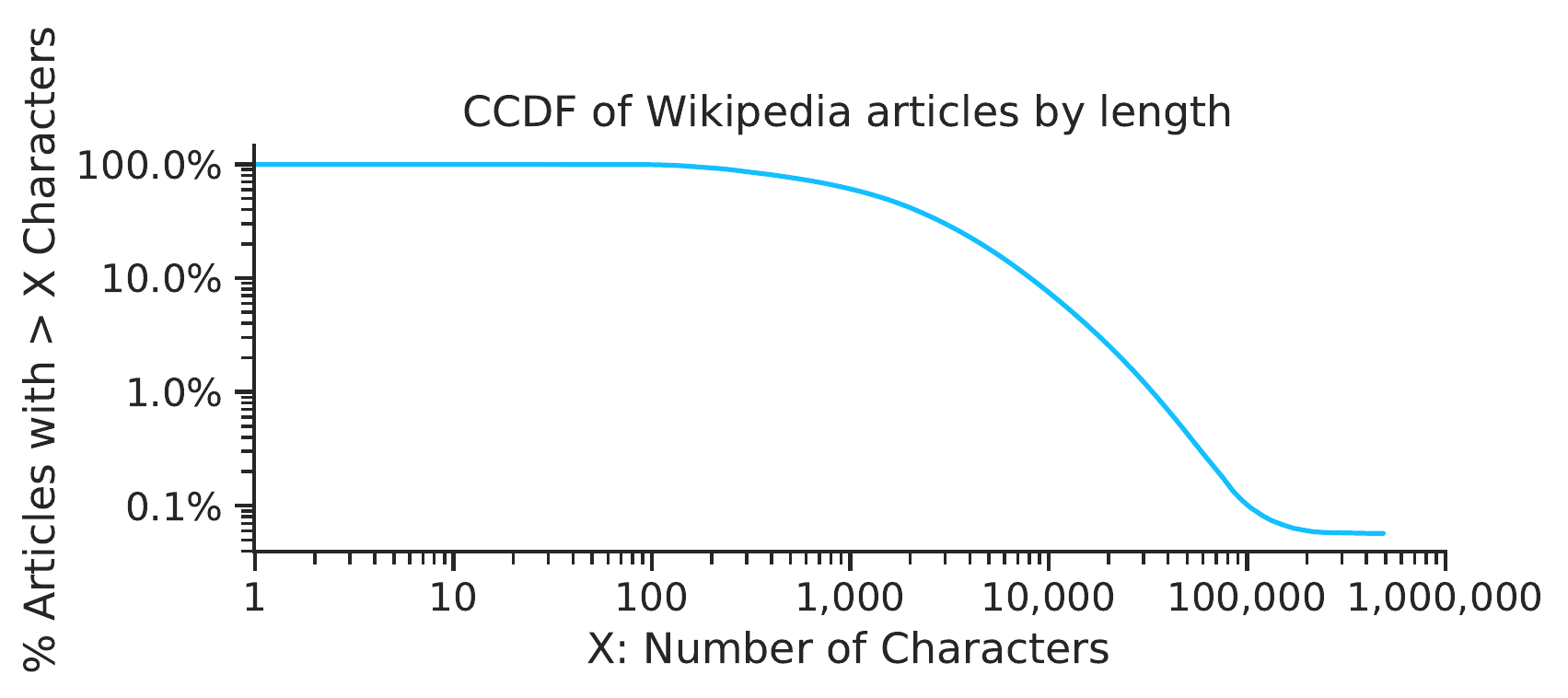}
        \caption{}
        \label{fig:length}
    \end{subfigure}
     \begin{subfigure}[b]{0.33\textwidth}
        \includegraphics[width=\textwidth,right]{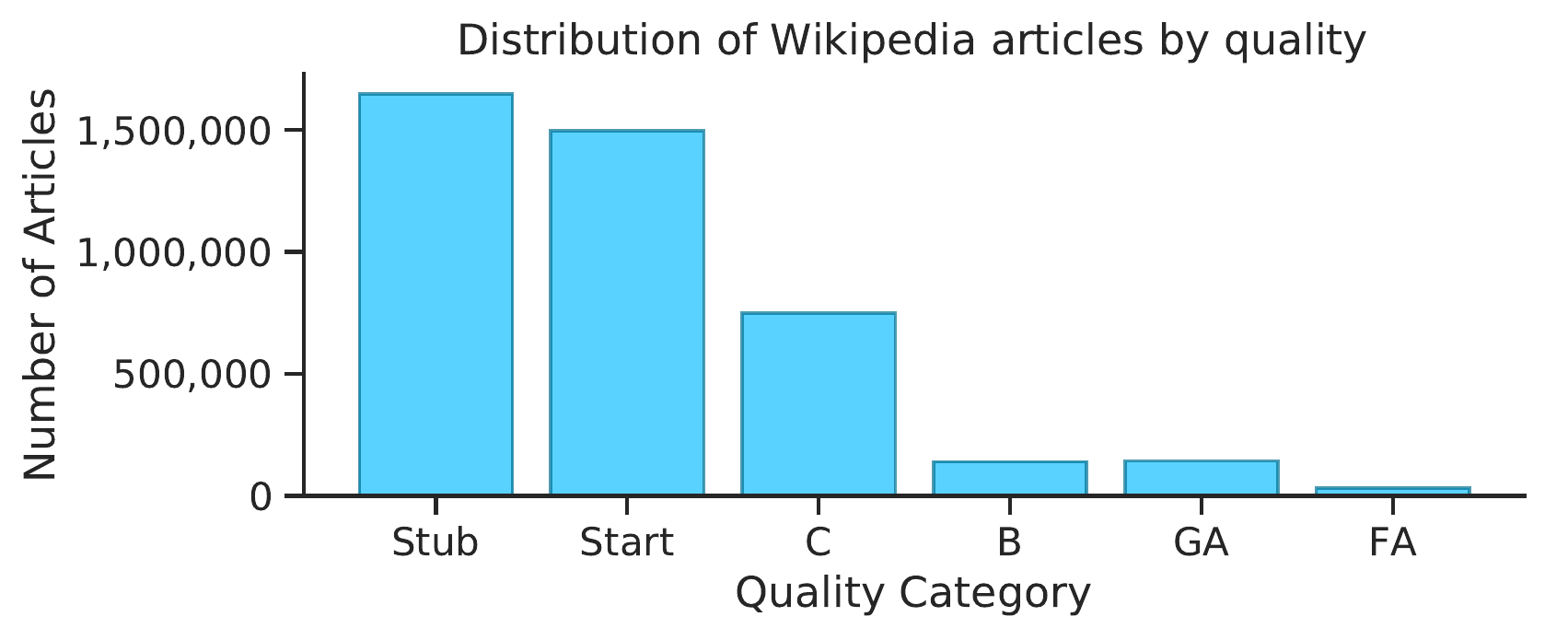}
        \caption{}
        \label{fig:quality}
    \end{subfigure}
\vspace{-15pt}
\caption{Distribution of Wikipedia articles by (a) popularity (number of pageviews), (b) page length (number of characters in wikicode), and (c) quality (increasing from left to right; ``GA'' for ``Good Article'', ``FA'' for ``Featured Article'') (\Secref{sec:General statistics of English Wikipedia}).}
\label{fig:article_stats}
 \end{figure*}

Throughout the paper, we will mostly focus on the data from the second data collection period (April 2019) and only use the October 2018 data for a longitudinal study measuring the impact of article quality on readers' engagement with citations.

\subsection{Definition of engagement metrics}
\label{sec:Definition of engagement metrics}

Two key metrics in our analysis will be the citation click-through rate (CTR) and the footnote hover rate.

For each page $p$ and each session $s$,
let $C(p,s)$ be the indicator function that is 1 if at least one reference was clicked on page $p$ during session $s$ by the respective user ($\refClick$ event), and 0 otherwise.
Analogously, let $H(p,s)$ indicate if the user hovered over at least one footnote ($\fnHover$ event).
Furthermore, let $N(p)$ be the number of sessions during which $p$ was loaded ($\pageLoad$ event)

\xhdr{Global click-through rate}
The global CTR measures overall reader engagement via reference clicks across Wikipedia.
It is defined as the fraction of page views on which at least one reference click occurred (treating all views of the same page in the same session as one single event):
\begin{equation}
\label{eqn:gctr}
\gctr = \frac{\sum_p \sum_s C(p,s)}{\sum_p N(p)},
\end{equation}
where $p$ ranges over the set of pages that contain at least one reference with a hyperlink.

\xhdr{Page-specific click-through rate}
The page-specific CTR for page $p$ is defined as the probability of observing at least one click on a reference in $p$ during a session in which $p$ was viewed:
\begin{equation}
\label{eqn:pctr}
\pctr(p) = \frac{\sum_s C(p,s)}{N(p)}.
\end{equation}

Finally, we denote the average page-specific CTR over a set $P$ of pages by
\begin{equation}
\label{eqn:avg_pctr}
\pctr(P) = \frac{1}{|P|} \sum_{p \in P} \pctr(p).
\end{equation}
Note that $\pctr(P)$ corresponds to a macro average where every page gets the same weight, whereas $\gctr$ corresponds to a micro average where pages are weighted in proportion to the number of sessions in which they were viewed.

\xhdr{Footnote hover rates}
In analogy to the above definitions, but when replacing the click indicator $C(p,s)$ with the hover indicator $H(p,s)$, we obtain the global and page-specific footnote hover rates:
\begin{equation}
\label{eqn:hover rate}
\ghr = \frac{\sum_p \sum_s H(p,s)}{\sum_p N(p)},
\hspace{5mm}
\phr(p) = \frac{\sum_s H(p,s)}{N(p)}.
\end{equation}

\subsection{Capturing event context}
\label{sec:Capturing event context}

Each event is characterized by a set of features that capture information about three aspects of the event: the session in which the event happened, the page, and the reference.

\begin{description}[leftmargin=*]
\item[Session:] We collect
the unique \textit{session token} (\cf\ \Secref{sec:Logging citation and page load events}) that identifies the
browser tab in which the event occurred.
\item[Pages:] At the article level, we store
\textit{title,} \textit{page id,}
\textit{text length of wikicode in characters,}
\textit{number of references,} and
\textit{popularity} (number of \pageLoad{} events during the data collection period).
We also use the ORES \textit{drafttopic} classifier ~\cite{Asthana:2018:FEH:3290265.3274290} to label each Wikipedia article with a vector of \textit{topics,} whose elements reflect the probability of the page to belong to one the 44 topics from the highest level of the  WikiProjects taxonomy.%
\footnote{\url{https://en.wikipedia.org/wiki/Wikipedia:WikiProject_Council/Directory}}
We further use the ORES \textit{articlequality} model ~\cite{halfaker_interpolating_2017} to label articles with a \textit{quality} level, which can take the following values (from low to high quality): ``Stub'', ``Start'', ``C-class'', ``B-class'', ``Good Article'', ``Featured Article''.
\item[References:] For each reference clicked or hovered, we record its \textit{URL,}
the \textit{text in the reference,} the \textit{text of the sentence} in which the reference is cited, and the \textit{relative position} (character offset from the start in plain text, divided by page length) in the page where the reference is cited. Since we associate references to their contexts, references to the same source appearing on different pages are treated as distinct. 
\end{description}

Wikipedia is dynamic by nature: articles are continuously updated, and their changes are tracked through revisions. To account for the evolution of articles over the 4 weeks of data collection, we aggregate individual revision-level metrics at the article level. To compute article-specific characteristics such as article length or number of references, we calculate their average over all revisions from the logging period.
To quantify the amount of reader engagement with a given article (\eg, page loads, reference clicks), we sum all events recorded at each revision of the article.

\subsection{General statistics of English Wikipedia}
\label{sec:General statistics of English Wikipedia}

By the end of the data collection, English Wikipedia contained 5.8M articles, 5.4M (95\%)
of which were loaded at least once in our data sample, in a total of 7.4M revisions.
Out of these articles, 3.9M (73\%) contain at least one citation, linking to a total of 24M distinct URLs.

Over the 4 weeks of data collection, we collected (at a 33\% sampling rate) 1.5B \pageLoad{} events (62\% from the mobile site and the rest from the desktop site).
In \Figref{fig:pageviews} we report the (complementary cumulative) popularity distribution for the Wikipedia pages that were viewed at least once during the data collection period.
The distribution is heavily skewed, with approximately 83\% of the articles loaded fewer than 100 times in the 33\% random sample (\cf \Secref{sec:Logging citation and page load events}), or fewer than 300 times when extrapolating to all data.

We observe a similar uneven distribution of page length (\Figref{fig:length}), with the majority of articles being very short.

\Figref{fig:quality} shows that the distribution of article quality levels is also heavily skewed toward low quality levels: most articles are identified as ``Stub'' or ``Start'', and fewer than 300K articles are marked as ``Good'' or ``Featured'' articles.

\begin{figure}[t]
\centering
\includegraphics[width=0.5\textwidth]{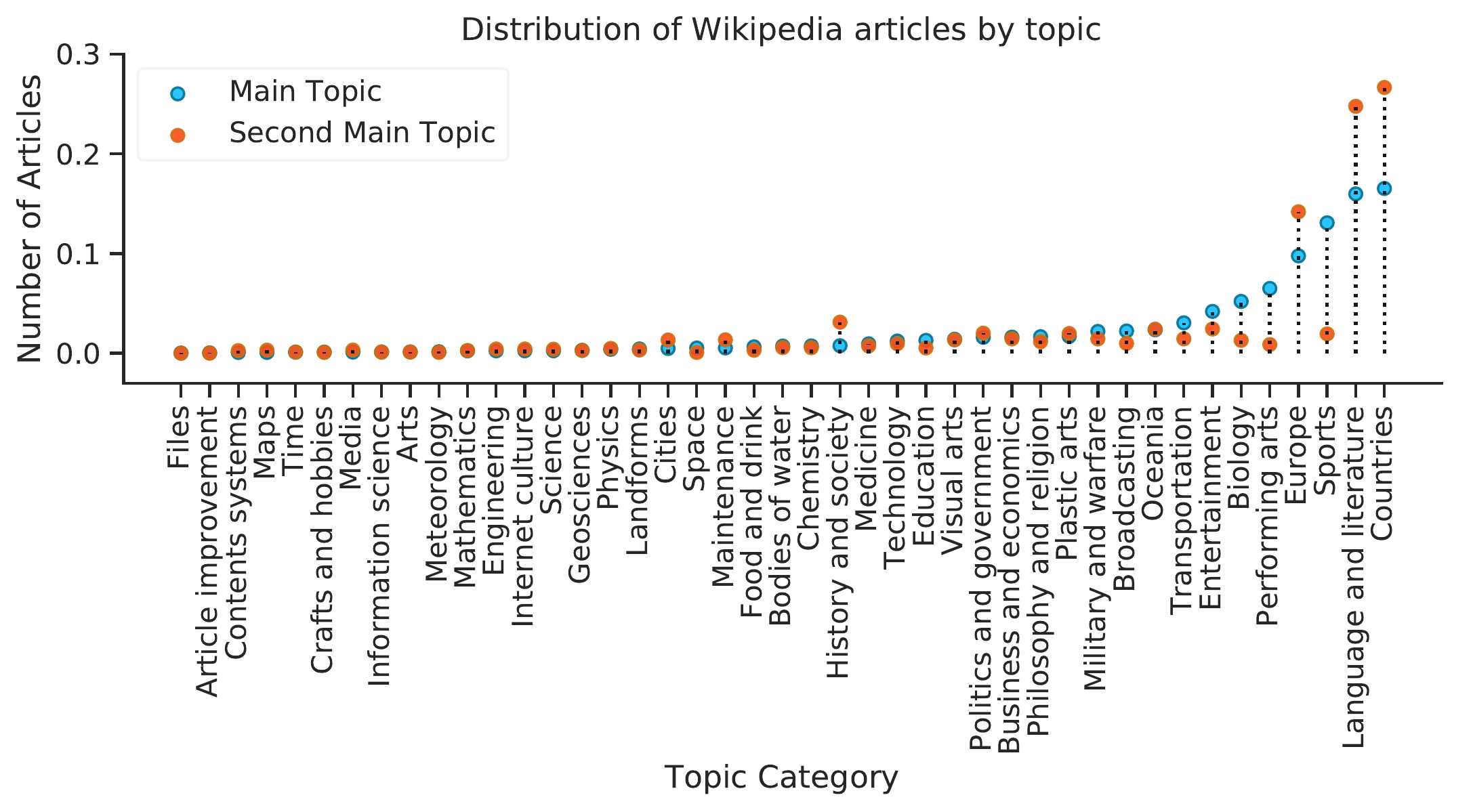}
\caption{Distribution of most and second most prominent Wikipedia article topics (\Secref{sec:General statistics of English Wikipedia}).}
\label{fig:topics}
\end{figure}

Finally (\Figref{fig:topics}), we find that a majority of articles are about geography or ``Language and literature'' (the latter including biographies), followed by topics related to sports and science.

%% file: 4_RQ1.tex
\section{RQ1: Prevalence of citation interactions}
\label{sec:RQ1}

After these preliminaries, we are now ready to address our first research question, which asks to what extent Wikipedia readers engage with citations.

\subsection{Distribution of interaction types}
\label{sec:Distribution of interaction types}

\begin{figure}[t]
    \centering
    \includegraphics[width=0.8\columnwidth]{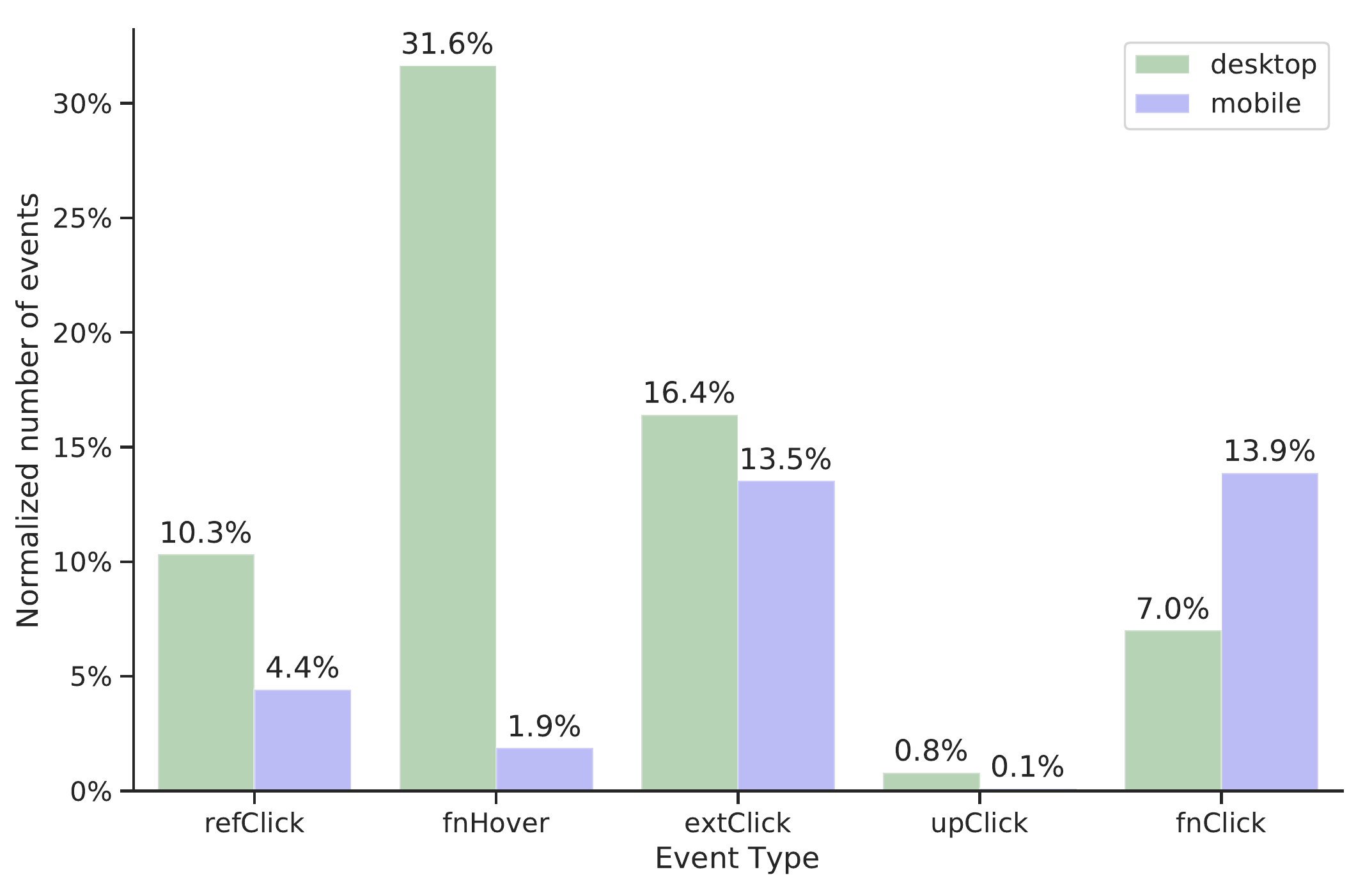}
    \caption{
    Relative frequency of citation-related events (\Secref{sec:Logging citation and page load events}), split into desktop (green, left bars) and mobile (blue, right bars) in April 2019 (\Secref{sec:Distribution of interaction types}).
    }
    \label{fig:event_distribution}
\end{figure}

We start by analyzing the relative frequency of the different citation events, as defined in \Secref{sec:Logging citation and page load events}.
Over the month of data collection, we captured a total of 96M citation events.
\Figref{fig:event_distribution} shows how these events distribute over the 5 event types, broken down by device type (mobile \vs{} desktop).
We observe that most interactions with citations happen on desktop rather than mobile devices, despite the fact that the majority of page loads (62\%) are made from mobile.

The interactions also distribute differently across types for mobile \vs\ desktop.
The by far prevailing event on desktop is hovering over a footnote (\fnHover) in order to display the reference text.
Hovering requires a mouse, which is not available on most mobile devices, which in turn explains the low incidence of \fnHover{} on mobile.
In order to reveal the reference text behind a footnote, mobile users instead need to click on the footnote, which presumably explains why \fnClick{} is the most common event on mobile.

Clicking external links outside of the \textit{References} section at the bottom of the page (\extClick) is the second most common event on both desktop and mobile, followed by clicks on citations from the \textit{References} section (\refClick). Finally, the \upClick{} action, which lets users jump back from the \textit{References} section to the locations where the citation is used in the main text, is almost never used.

\subsection{Citation click-through rates}
\label{sec:Citation click-through rates}

We now focus on the two prevalent interactions with citations, hovering over footnotes (\fnHover) and leaving Wikipedia by clicking on citation links (\refClick).
(We do not dwell on \extClick{} events, as they do not concern citations but other external links; \cf \Secref{sec:Logging citation and page load events}.)

First, we observe that, out of the 24M distinct URLs that are cited across all articles in English Wikipedia,
93\% of the URLs are never clicked during our month of data collection.

Next, we note that the global click-through rate (CTR) across all pages with at least one citation (\gctr, \Eqnref{eqn:gctr}) is 0.29\%; \ie, clicks on references happen on fewer than 1 in 300 page loads.
Breaking the analysis up by device type, we observe again substantial differences between desktop and mobile: on desktop the global CTR is 0.56\%, over 4 times as high as on mobile, where it is only 0.13\%.

The average page-specific CTR (\pctr, \Eqnref{eqn:avg_pctr}) is higher, at 1.1\% for desktop and 0.52\% for mobile. This is due to the fact that there are many rarely viewed pages (\cf\ \Figref{fig:pageviews}) with a noisy, high CTR.
After excluding pages with fewer than 100 page views, the global CTR is 0.67\% on desktop, and 0.21\% on mobile.

Engagement via footnote hovering is slightly higher, at a global footnote hover rate (\ghr, \Eqnref{eqn:hover rate}) of 1.4\%.
The average page-specific footnote hover rate (\phr, \Eqnref{eqn:hover rate}) is 0.68\% when including all pages with at least one clickable reference, and 1.1\% when excluding pages with fewer than 100 page views.\footnote{As mentioned in \Secref{sec:Distribution of interaction types}, hovering is not available on most mobile devices, so the hovering numbers pertain to desktop devices only.}

Given these numbers, we conclude that readers' engagement with citations is overall low.

\subsection{Positional bias}
\label{sec:Positional bias}

\begin{figure}[t]
    \centering
    \includegraphics[width=0.8\columnwidth]{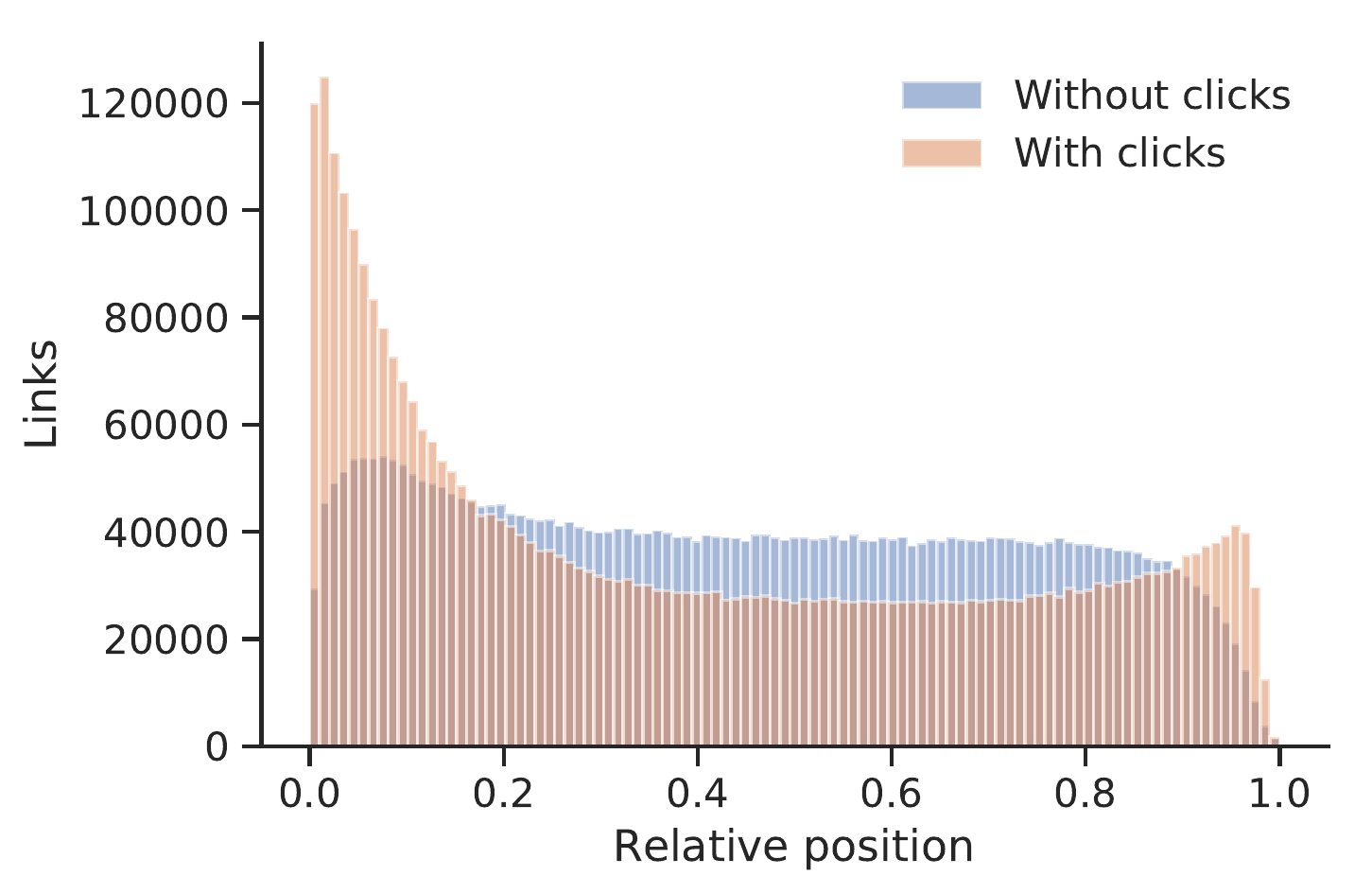}
    \caption{Relative position in page of clicked \vs\ unclicked references, for references with hyperlinks (\Secref{sec:Positional bias}).}
    \label{fig:click_pos}
\end{figure}

Previous work has shown that users are more likely to click Wikipedia\hyp internal links that appear at the top of a page \cite{paranjape_improving_2016}.
To verify whether this also holds true for references, we sample one random page load with citation interactions per session and randomly sample one clicked and one unclicked reference for this page load.
We then compute each reference's relative position in the page as the offset from the top of the page divided by the page length (in characters).
\Figref{fig:click_pos}, which shows the distribution of the relative position for clicked and unclicked references, reveals that users are more likely to click on references toward the top and (less extremely so) the bottom of the page.

\subsection{Top clicked domains}
\label{sec:Frequency of clicked domains}

Next, we investigate what are the most frequent domains at which users arrive upon clicking a citation.

Initially, we found that the most frequently clicked domain is \texttt{archive.org} (Internet Archive), with
882K \refClick{} events.
Such URLs are usually snapshots of old Web pages archived by the Internet Archive's Wayback Machine.
To handle such cases, we extract the original source domains from wrapping \texttt{archive.org} URLs.

\begin{figure*}[t]
    \begin{subfigure}[b]{0.42\textwidth}
        \includegraphics[width=\textwidth,left]{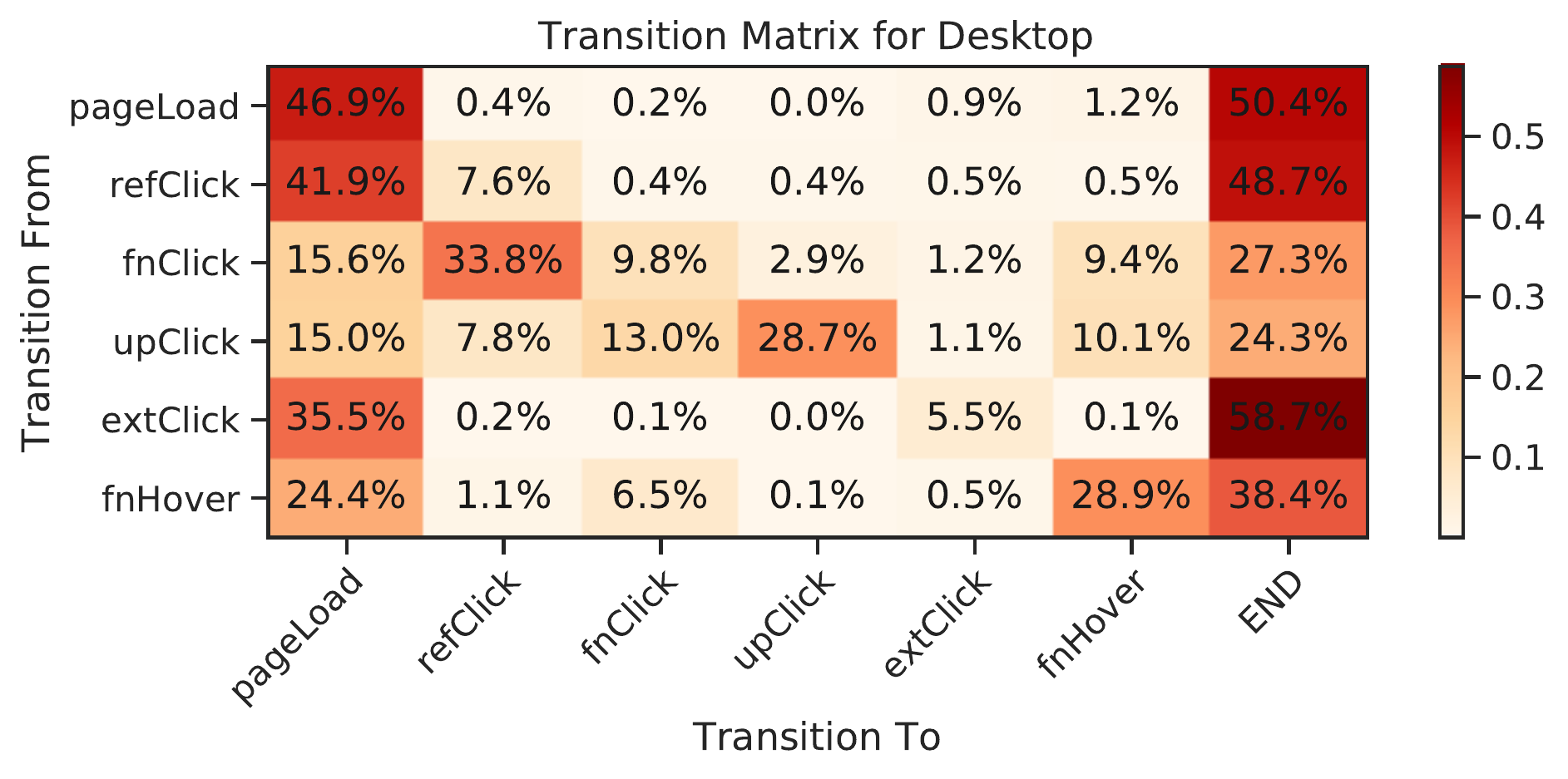}
        \caption{Reader behavior on desktop devices.}
        \label{fig:desktoptransitions}
    \end{subfigure}
   \begin{subfigure}[b]{0.42\textwidth}
        \includegraphics[width=\textwidth,right]{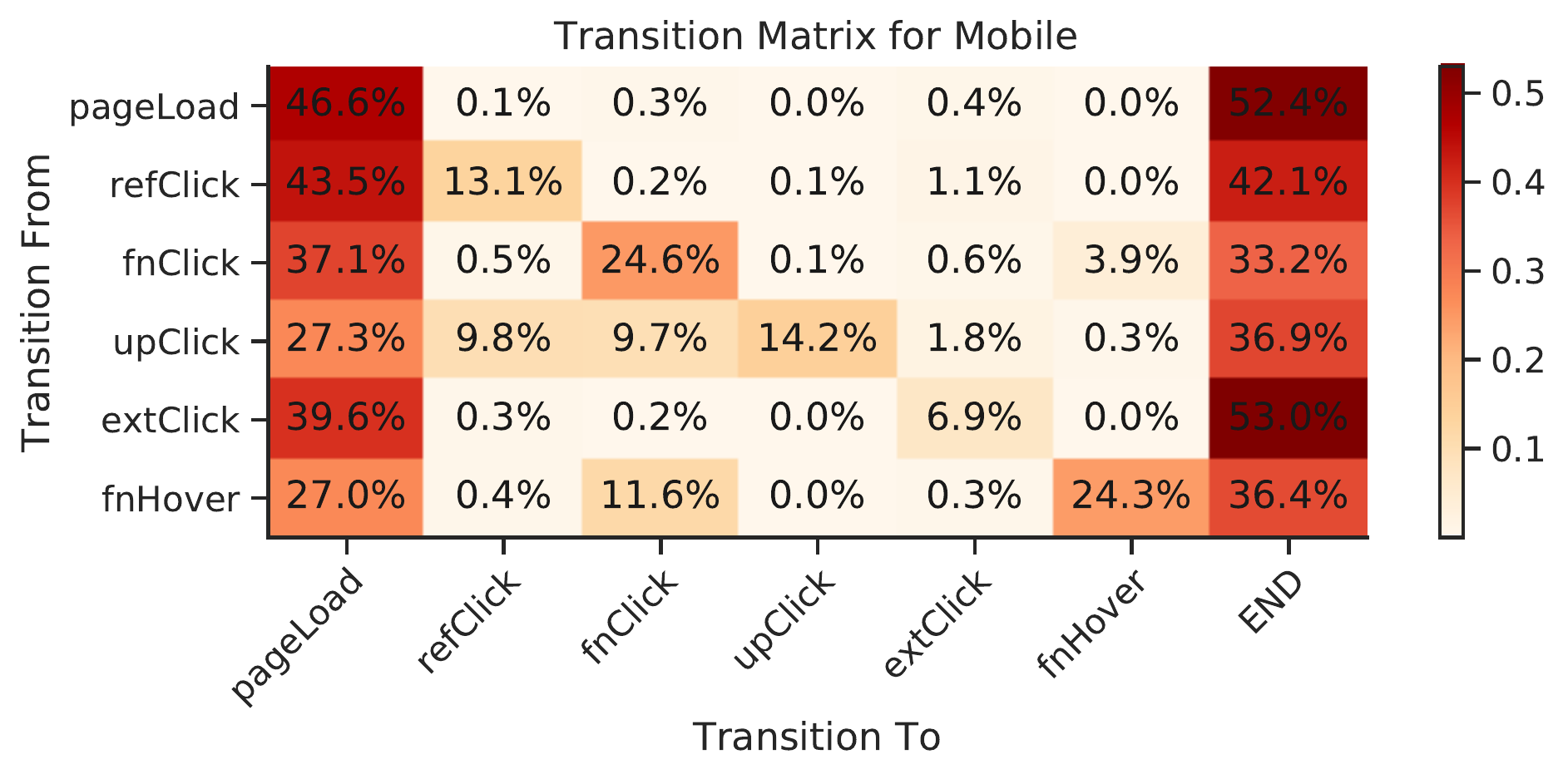}
        \caption{Reader behavior on mobile devices.}
        \label{fig:mobiletransitions}
    \end{subfigure}
    \caption{
    Transition matrices of first-order Markov chains for (a) desktop devices and (b) mobile devices, aggregating reader behavior with respect to citation events when navigating a Wikipedia article with references (\Secref{sec:Markovian analysis of citation interactions}).}
    \label{fig:transitions}
\end{figure*}

\begin{figure}[t]
\centering
\includegraphics[width=0.8\columnwidth]{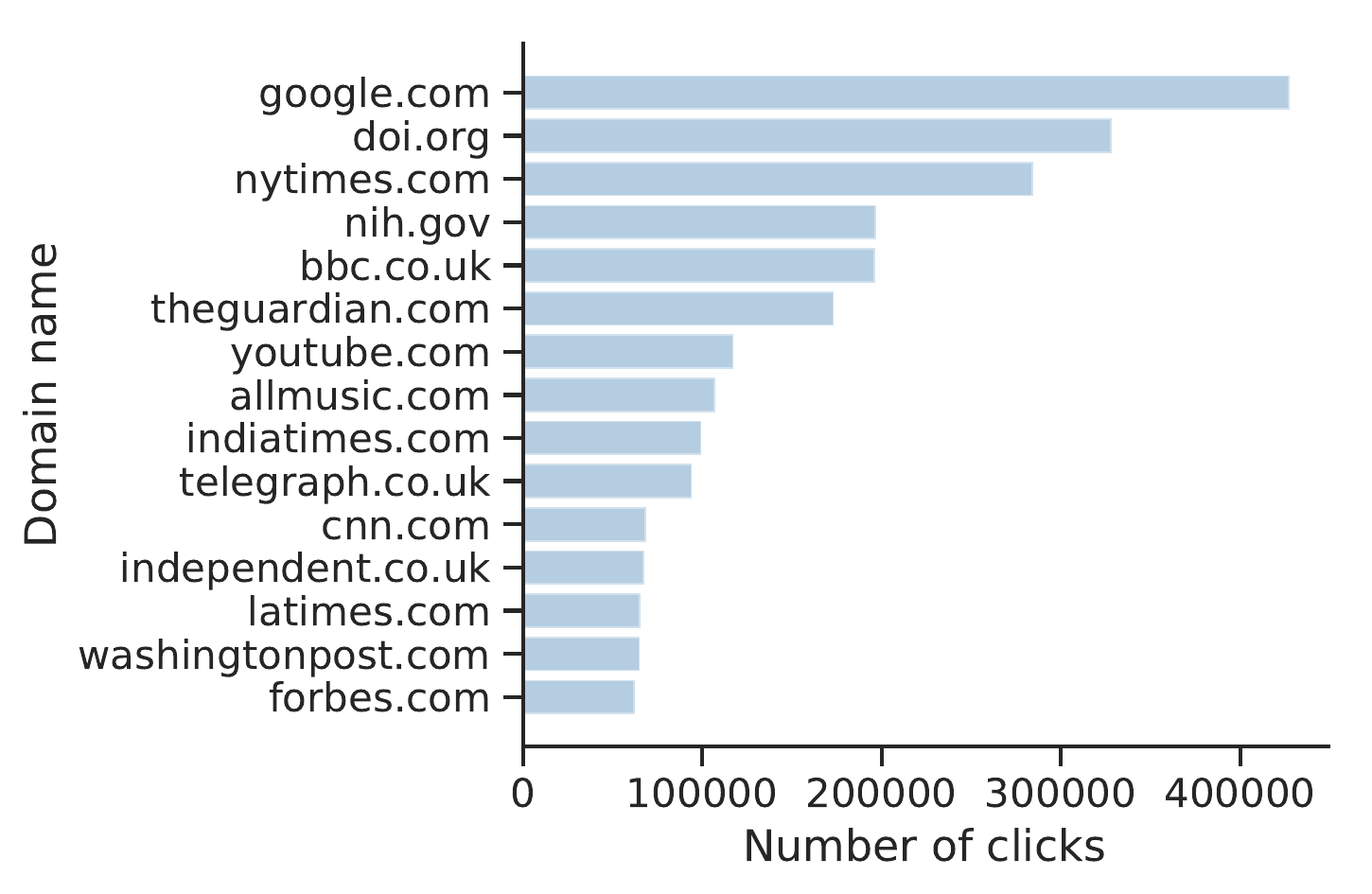}
\caption{Top 15 domain names appearing in English Wikipedia references (\Secref{sec:Frequency of clicked domains}), sorted by number of clicks received during April 2019.}
\label{fig:domains}
\end{figure}

In \Figref{fig:domains} we report the top 15 domains by number of \refClick{} events. The most clicked domain
is \texttt{google.com}.
Drilling deeper, we checked the main subdomains contributing to this statistic, finding that a significant proportion of clicks goes to \texttt{books.google.com}, which is providing partial access to printed sources. The second most clicked domain is \texttt{doi.org}, the domain for all scholarly articles, reports, and datasets recorded with a Digital Object Identifier (DOI), followed by (mostly liberal) newspapers (\textit{The New York Times}, \textit{The Guardian}, \etc)\ and broadcasting channels (BBC).

\subsection{Markovian analysis of citation interactions}
\label{sec:Markovian analysis of citation interactions}

Whereas the above analyses involved individual events, we now begin to look at \textit{sessions:} sequences of events that occurred in the same browser tab (as indicated by the session token; \Secref{sec:Logging citation and page load events}).
Every session starts with a \pageLoad{} event, and we append a special \END{} event after the last actual event in each session.

By counting event transitions within sessions, we construct the first-order Markov chain that specifies the probability $P(j|i)$ of observing event $j$ right after event $i$, where $i$ and $j$ can take values from the event set introduced in \Secref{sec:Logging citation and page load events} (\pageLoad, \refClick, \extClick, \fnClick, \upClick, \fnHover) plus the special \END{} event.

The transition probabilities are reported in \Figref{fig:transitions}. We observe that most reading sessions are made up of page views only: on both desktop and mobile, after loading a page, readers tend to end the session (with a probability of around 50\%) or load another page in the same tab (47\%).
All citation\hyp related events have a very low probability (at most 1.2\%) of occurring right after loading a page.

On desktop, reference clicks become much more likely after footnote clicks (34\%), and footnote clicks in turn become much more likely after footnote hovers (6.5\%), hinting at a common 3-step motif (\fnHover, \fnClick, \refClick), where the reader engages ever more deeply with the citation.
Note, however, that this is not true for mobile devices, where, even after readers clicked on a footnote, the probability of also clicking on the citation stays low~(0.5\%).

Finally, reference clicks (\refClick) are also common immediately after other reference clicks (8\% on desktop, 13\% on mobile).
Note that for external links outside of the \textit{References} section (\extClick) we see a different picture:
such external clicks are only rarely followed by interactions with citations (\fnHover, \fnClick, \refClick),
and in the majority of cases (59\% on desktop, 53\% on mobile) they conclude the session, suggesting that Wikipedia is in these cases commonly used as a gateway to external websites.

%% file: 5_RQ2.tex
\section{RQ2: Page-level analysis of citation interactions}
\label{sec:RQ2}

We now proceed to our second research question, which asks what features of a Wikipedia page predict whether readers will engage with the references it contains.

\begin{figure}[t]
    \centering
    \includegraphics[width=0.9\columnwidth]{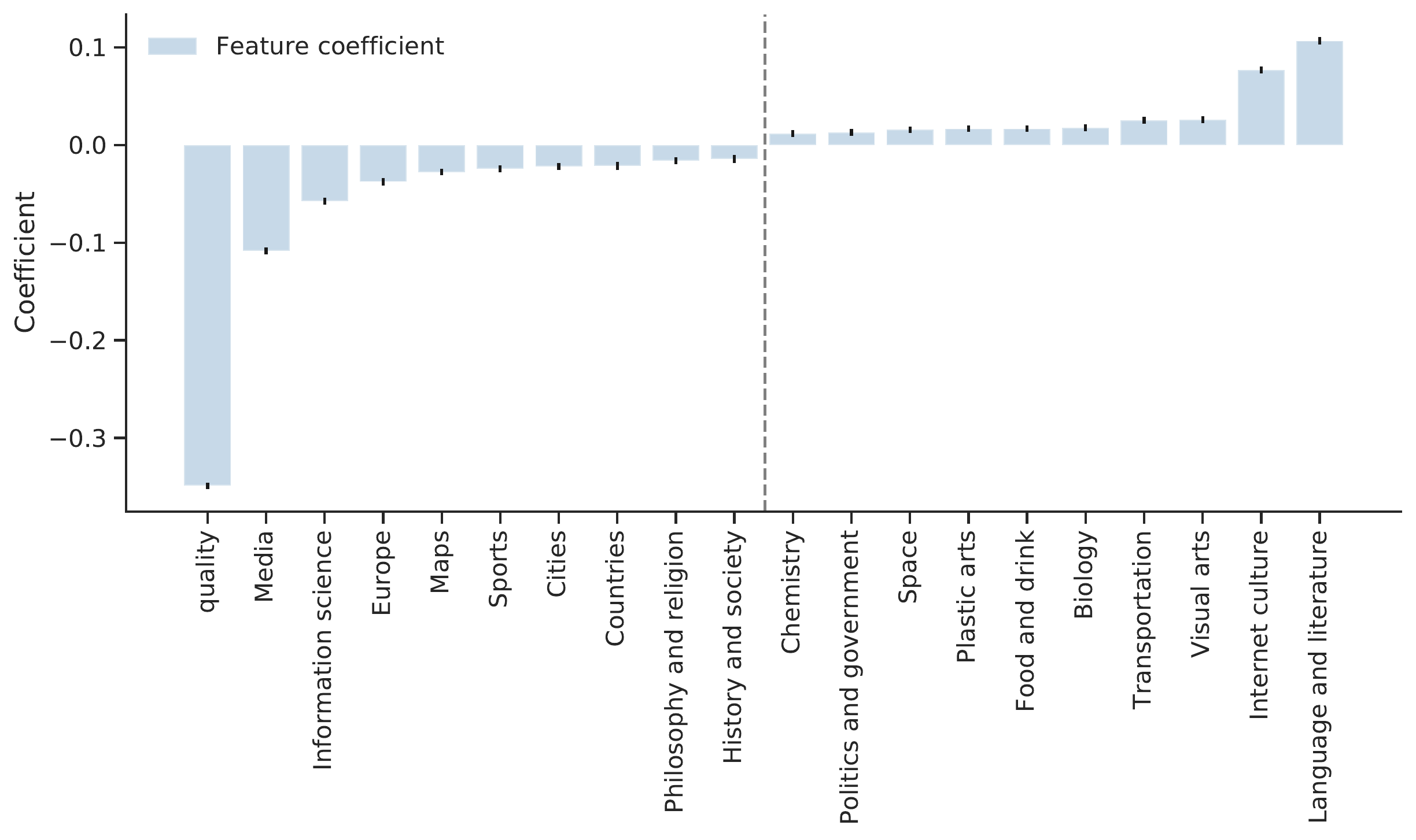}
    \caption{Contribution of features to logistic regression model predicting if \refClick{} event will eventually occur after page load (\Secref{sec:Predictors of reference clicks}). Top 10 positive and negative coefficients shown, with 95\% CIs.}
    \label{fig:features_contribution}
\end{figure}

\subsection{Predictors of reference clicks}
\label{sec:Predictors of reference clicks}

As a first step, we perform a regression analysis.
We train a logistic regression classifier for predicting whether a given \pageLoad{} event will eventually be followed by a \refClick{} event.
To assemble the training set, we first find sessions with 
at least one (positive) \pageLoad{} followed by a \refClick{} and
at least one (negative) \pageLoad{} not followed by a \refClick{},
and make sure to include at most one such pair per session in order to avoid over\hyp representing power users with extensive sessions.
The dataset totals
938K 
pairs, which we split into 80\% for training and 20\% for testing.

As predictors we use the article's \textit{topic} vector (with entries from $[0,1]$; \Secref{sec:Capturing event context}) and the \textit{quality} label (\Secref{sec:Capturing event context}), which we also normalize to a score in the range $[0,1]$ using the mapping from a previous study \cite{halfaker_interpolating_2017}.
We did not use the number of references and the length of the page, as they are important features in the quality model and would cause collinearity issues due to their high correlation with quality (Pearson's correlation 0.81
and 0.75,
respectively).

The resulting regression model has an area under the ROC curve (AUC) of 0.6 on the testing set.
A summary of the 10 most predictive positive and negative coefficients is given in \Figref{fig:features_contribution}.
By far the most important predictor---with a large negative weight---is the article's quality.
Moreover, some topics are positive predictors (\eg, ``Language and literature'', which also includes all biographies, as well as ``Internet culture''), while others are negative predictors (\eg, ``Media'', ``Information science'').

Given the importance of the quality feature in this first analysis, we now move to investigating its role in a more controlled study.

\subsection{Effects of page quality}
\label{sec:Effects of page quality}
\label{sec:ArticleQualityExperiment}

To come closer to a causal understanding of the impact of an article's quality on readers' clicking citations in the article, we perform a matched observational study.
The ideal goal would be to compare the page\hyp specific CTR (\Eqnref{eqn:pctr}) for pairs of articles---one of high, the other of low quality---that are identical in all other aspects.

\begin{figure}[t]
    \centering
    \includegraphics[width=0.8\columnwidth]{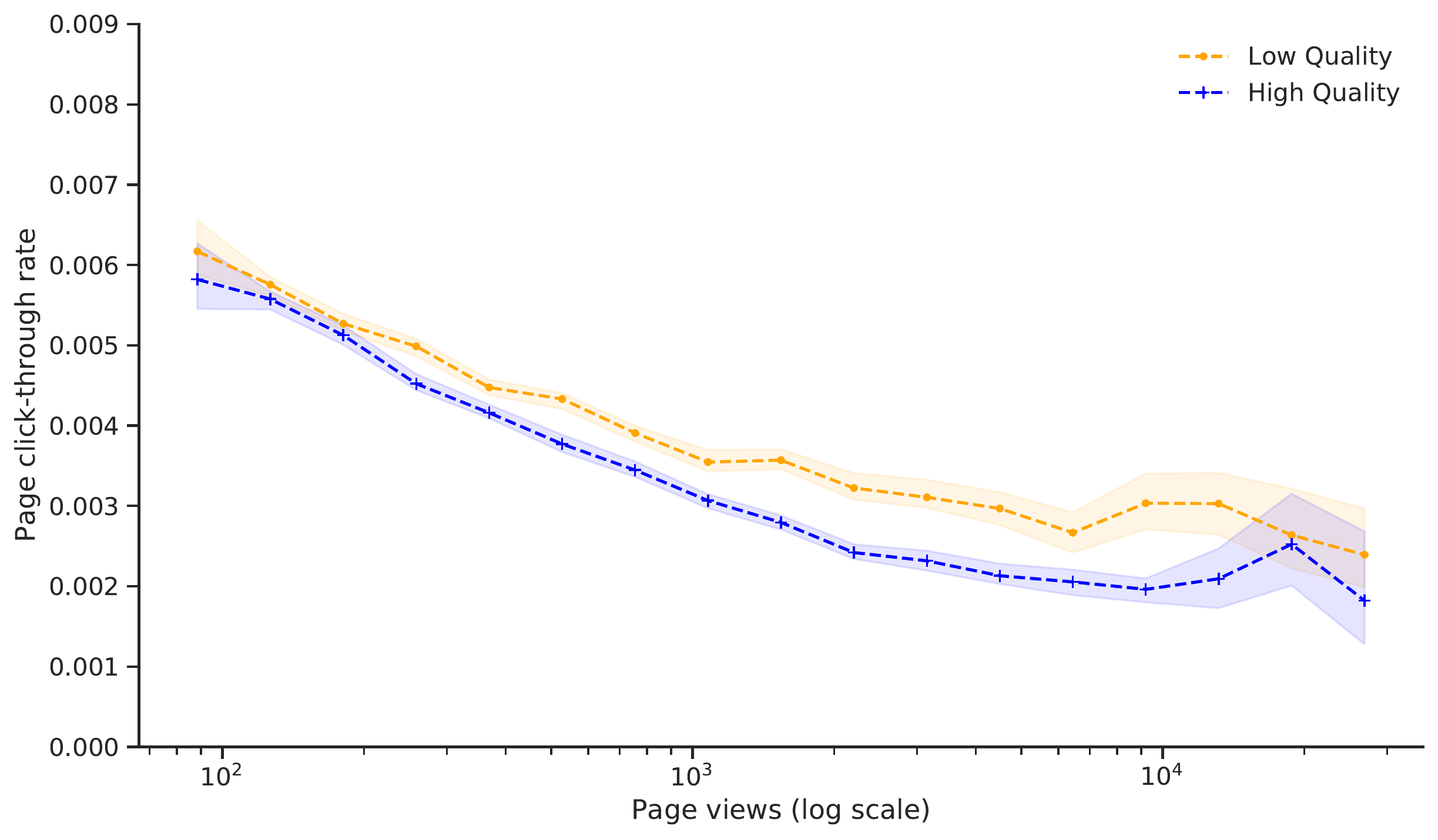}
    \caption{Comparison of page-specific click-through rate for low- (yellow) \vs\ high\hyp quality (blue) articles, as function of popularity (\Secref{sec:Effects of page quality}). Error bands: bootstrapped 95\% CIs.}
    \label{fig:quality_ct_rate}
\end{figure}

\xhdr{Propensity score}
Finding such exact matches is unrealistic in practice, so we resort to propensity score matching~\cite{austin2011introduction}, which provides a viable solution.
The propensity score specifies the probability of being treated as a function of the observed (pre-treatment) covariates.
Crucially, data points with equal propensity scores have the same distribution over the observed covariates, so matching treated to untreated points based on propensity scores will balance the distribution of observed covariates across treatment groups.

In our setting, we define being of high quality as the treatment and estimate propensity scores via a logistic regression that uses topics, length, number of citations, and popularity as observed covariates in order to predict quality as the binary treatment variable.
We consider as low-quality all articles tagged as \textit{Stub} or \textit{Start} (74\% of the total; \Figref{fig:quality}), and as high-quality the rest. 
Articles without a \refClick{} or fewer than 100 \pageLoad{} events are discarded in order to avoid noisy estimates of the page\hyp specific CTR.
This leaves us with 854K articles.

\xhdr{Matching}
We compute a matching (comprising 198K pairs) that minimizes the total absolute difference of within-pair propensity scores, under the constraint that the length of matched pages should not differ by more than 10\%.
This constraint is necessary to ascertain balance on the page length feature because page length is so highly correlated with quality (Pearson correlation 0.81; \cf\ \Secref{sec:Predictors of reference clicks}).
After matching, we manually verify that all observed covariates, including page length, are balanced across groups.

\xhdr{Results}
\Figref{fig:quality_ct_rate} visualizes the average page\hyp specific CTR for articles of low (yellow) and high (blue) quality as a function of article popularity.
We can observe that the CTR of low-quality articles significantly surpasses that of high-quality articles across all levels of popularity.
In interpreting this result, it is important to recall that page length is one of the most important features in ORES \cite{halfaker_interpolating_2017}, the quality\hyp scoring model we use here.
As we control for page length, the gap observed in \Figref{fig:quality_ct_rate} may be attributed to the remaining features used by ORES, such as the presence of an infobox, the number of images, and the number of sections and subsections.

We hence dedicate our next, final page-level analysis to estimating the impact of page length alone on page\hyp specific CTR.

\begin{figure}[t]
    \centering
    \includegraphics[width=0.8\columnwidth]{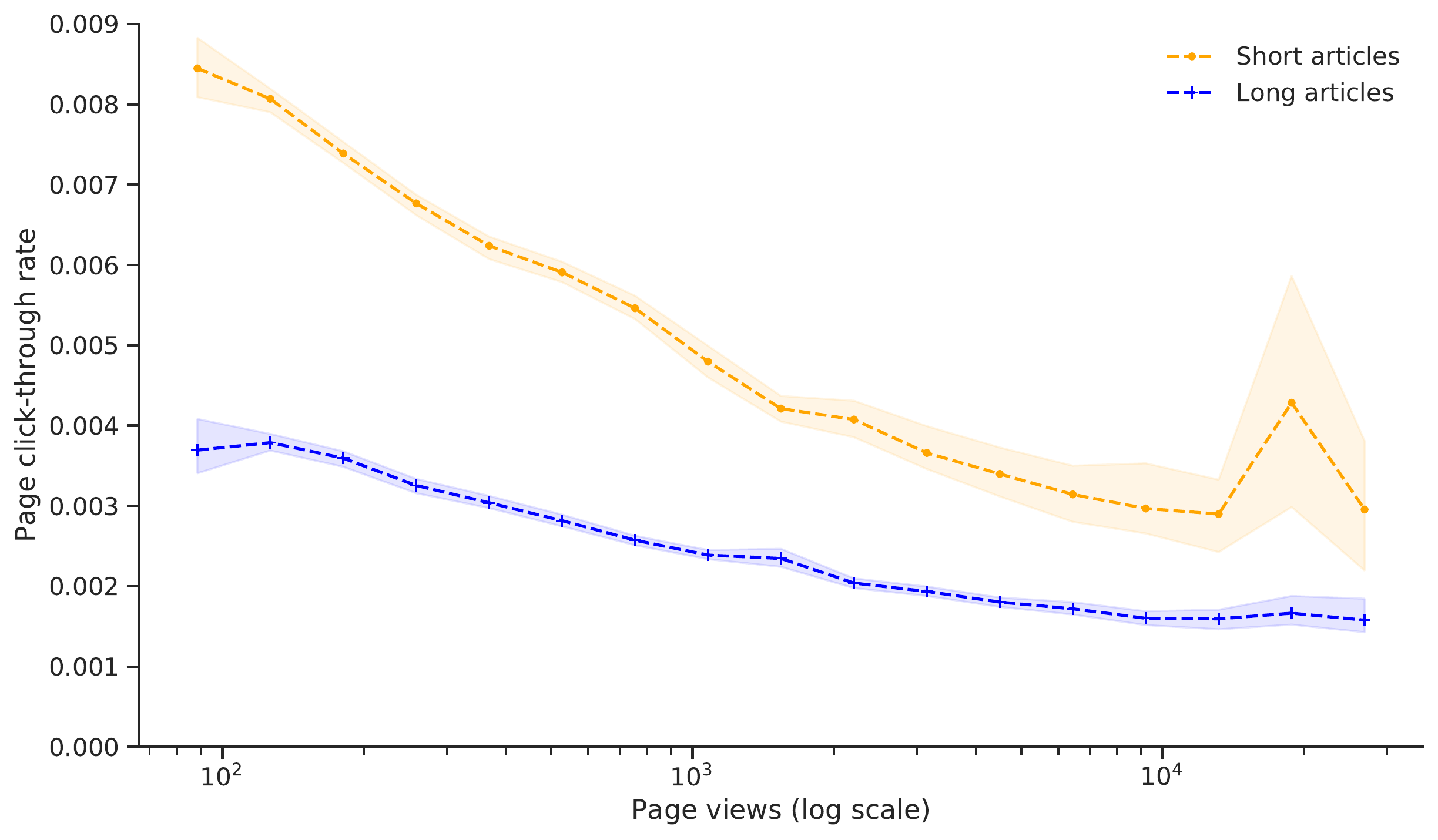}
    \caption{
    Comparison of page-specific click-through rate for short (yellow) \vs\ long (blue) articles, as function of popularity (\Secref{sec:Effects of page length}). Error bands: bootstrapped 95\% CIs.
    }
    \label{fig:length_ct_rate}
    \vspace{-5pt}
\end{figure}

\subsection{Effects of page length}
\label{sec:Effects of page length}

In order to measure the effect of page length on CTR, we take a two-pronged approach, first via a cross\hyp sectional study using propensity scores, and second via a longitudinal study.

\xhdr{Cross-sectional study}
First, we conduct a matched study based on propensity scores analogous to \Secref{sec:Effects of page quality}, but now with page length as the treatment variable (using the longest and the shortest 40\% of articles as treatment groups),
and all other features (except quality) as observed covariates.
Matching yields 683K pairs, and we again manually verify covariate balance across treatment groups.

The average page\hyp specific CTR of short articles (0.68\%) is more than double that of long articles (0.27\%; $p \ll 0.001$ in a two-tailed Mann--Whitney $U$ test).
Moreover, as seen in \Figref{fig:length_ct_rate}, this relative difference obtains across all levels of article popularity.

\begin{figure}[t]
    \centering
    \includegraphics[width=0.8\columnwidth]{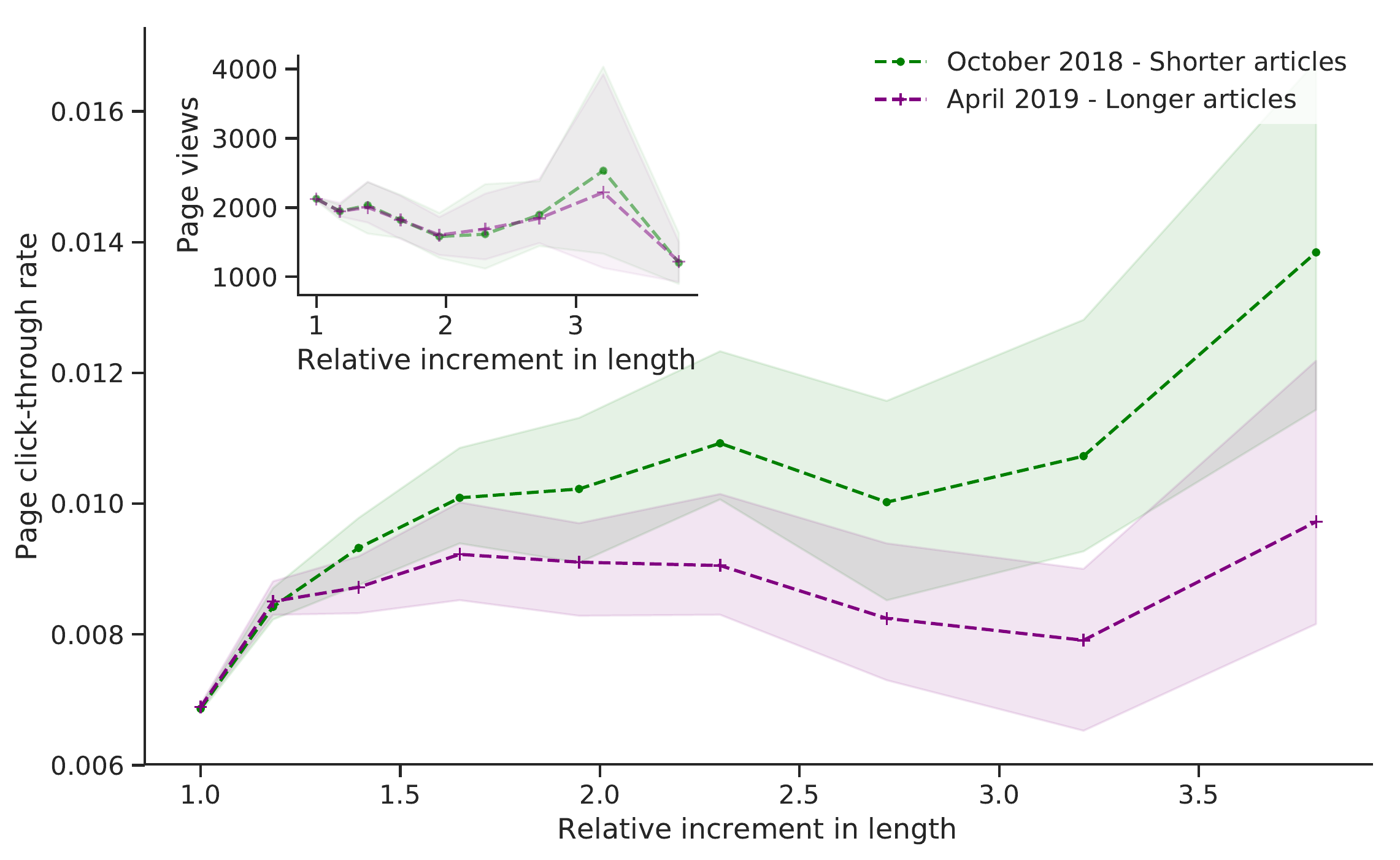}
    \caption{Comparison of page-specific click-through rate of shorter (green) \vs\ longer (purple) revisions of identical articles, as function of length ratio (\Secref{sec:Effects of page length}).
    Inset: popularity as function of length ratio. Error bands: bootstrapped 95\% CIs.}
    \label{fig:time_diff}
\end{figure}

\xhdr{Longitudinal study}
While in the above cross-sectional study propensity score matching ensures that the covariates of long \vs\ short articles are indistinguishable at the aggregate treatment group level, it does not necessarily do so at the pair level.
Also, we did not include as observed covariates features describing the users who read the respective articles, and it might indeed be the case that users with a liking for short, niche articles also have a higher probability of clicking citations.
In order to mitigate the danger of such remaining potential confounds and achieve even finer control, we now conduct a longitudinal study to assess how a variation in length of the \textit{same} article impacts its CTR.

To do so, we select all articles that grew in length between October 2018 and April 2019, our two data collection periods (\Secref{sec:Logging citation and page load events}).
To control for the effect of page popularity, which was observed to negatively correlate with CTR (\Figref{fig:quality_ct_rate} and \ref{fig:length_ct_rate}), we assign a popularity level to each article by binning page view counts into deciles and discard articles whose popularity level has changed between the two periods.
This way, we obtain a set of 120K articles with matched long and short revisions.

By grouping these articles by the length ratio of their two revisions and plotting this ratio against the CTR for the long (purple) \vs\ short (green) versions (\Figref{fig:time_diff}), we provide a further strong indicator that page length causally decreases the prevalence of citation clicking.
According to a Mann--Whitney $U$ test, the CTR difference between long and short revisions is statistically significant with $p < 0.05$ starting from a length increase of 17\%, and with $p < 0.01$ from 31\%.
In addition, to verify that the effect is not confounded by a concomitant change in article popularity, the inset plot in \Figref{fig:time_diff} shows that the popularity indeed stays constant between revisions.

\begin{table}
\tiny
\begin{tabular}{c | lr | lr | lr | lr} 
&
\multicolumn{4}{c|}{Positive contribution} &
\multicolumn{4}{c}{Negative contribution} \\

\hline

&
\multicolumn{2}{c|}{In sentence}&
\multicolumn{2}{c|}{In reference} &
\multicolumn{2}{c|}{In sentence}&
\multicolumn{2}{c}{In reference} \\
\hline

&
Word & Coeff. & Word & Coeff. & Word & Coeff. & Word & Coeff. \\
\hline
 \parbox[t]{2mm}{\multirow{10}{*}{\rotatebox[origin=c]{90}{All topics}}}

& greatest & 0.36 & know & 0.25 & debut & -0.25 & awards & -0.33\\
& born & 0.28 & pmc & 0.24 & moved & -0.16 & deadline & -0.32\\
& died & 0.23 & 2019 & 0.21 & worked & -0.16 & billboard & -0.17\\
& website & 0.23 & website & 0.21 & awarded & -0.16 & register & -0.17\\
& ranked & 0.23 & dies & 0.20 & joined & -0.13 & link & -0.16\\
& known & 0.20 & former & 0.19 & began & -0.13 & isbn & -0.15\\
& professional & 0.19 & family & 0.16 & appeared & -0.12 & board & -0.14\\
& relationship & 0.19 & behind & 0.15 & score & -0.11 & variety & -0.14\\
& rating & 0.18 & allmusic & 0.15 & festival & -0.11 & next & -0.14\\
& article & 0.18 & story & 0.15 & attended & -0.11 & archive & -0.13\\
\hline

\hline

 \parbox[t]{2mm}{\multirow{10}{*}{\rotatebox[origin=c]{90}{STEM}}}

& online & 0.25 & definition & 0.30 & requirements & -0.17 & oclc & -0.26\\
& tests & 0.23 & 2019 & 0.24 & run & -0.17 & best & -0.23\\
& 2019 & 0.23 & free & 0.22 & rather & -0.16 & jstor & -0.22\\
& short & 0.17 & pmc & 0.21 & another & -0.15 & evaluation & -0.16\\
& known & 0.17 & website & 0.20 & said & -0.15 & wiley & -0.16\\
& algorithms & 0.16 & pdf & 0.19 & launched & -0.15 & london & -0.15\\
& published & 0.16 & overview & 0.17 & less & -0.14 & isbn & -0.14\\
& defined & 0.15 & methods & 0.15 & make & -0.12 & internet & -0.14\\
& programming & 0.15 & introduction & 0.14 & better & -0.12 & industrial & -0.14\\
& digital & 0.15 & years & 0.13 & popular & -0.12 & source & -0.14\\

\hline

 \parbox[t]{2mm}{\multirow{10}{*}{\rotatebox[origin=c]{90}{Culture}}}
& article & 0.30 & daughter & 0.36 & indicating & -0.42 & awards & -0.36\\
& born & 0.28 & obituary & 0.31 & premiered & -0.28 & award & -0.33\\
& greatest & 0.27 & know & 0.31 & chart & -0.21 & deadline & -0.28\\
& professional & 0.27 & instagram & 0.29 & debut & -0.21 & cast & -0.22\\
& died & 0.26 & boy & 0.28 & moved & -0.20 & global & -0.21\\
& known & 0.25 & sex & 0.25 & began & -0.17 & next & -0.19\\
& ranked & 0.24 & wife & 0.24 & earned & -0.16 & isbn & -0.18\\
& relationship & 0.23 & former & 0.24 & recorded & -0.16 & drama & -0.18\\
& website & 0.23 & historic & 0.24 & alongside & -0.16 & standard & -0.18\\
& sexual & 0.23 & 2019 & 0.23 & worked & -0.16 & tour & -0.18\\

\hline

 \parbox[t]{2mm}{\multirow{10}{*}{\rotatebox[origin=c]{90}{History and Society}}}
& born & 0.29 & definition & 0.43 & came & -0.20 & jstor & -0.25\\
& website & 0.21 & overview & 0.22 & award & -0.16 & record & -0.21\\
& 2019 & 0.21 & best & 0.19 & transportation & -0.13 & link & -0.20\\
& died & 0.20 & 2019 & 0.19 & protection & -0.12 & 2002 & -0.17\\
& currently & 0.19 & website & 0.19 & member & -0.12 & election & -0.16\\
& known & 0.17 & statistics & 0.17 & began & -0.11 & 1998 & -0.15\\
& referred & 0.17 & death & 0.16 & originally & -0.11 & ed & -0.15\\
& customers & 0.16 & last & 0.16 & specific & -0.11 & isbn & -0.15\\
& study & 0.16 & ship & 0.15 & awarded & -0.10 & announces & -0.14\\
& activities & 0.15 & top & 0.15 & addition & -0.10 & board & -0.12\\

\hline

 \parbox[t]{2mm}{\multirow{10}{*}{\rotatebox[origin=c]{90}{Geography}}}
& politician & 0.50 & woman & 0.34 & debut & -0.45 & crime & -0.28\\
& born & 0.26 & know & 0.27 & missing & -0.22 & awards & -0.28\\
& magazine & 0.25 & dies & 0.26 & career & -0.21 & register & -0.24\\
& believed & 0.23 & family & 0.23 & timmothy & -0.20 & link & -0.24\\
& married & 0.23 & website & 0.20 & executive & -0.19 & interview & -0.19\\
& ranked & 0.22 & mail & 0.19 & episode & -0.17 & 2000 & -0.17\\
& video & 0.22 & father & 0.18 & months & -0.17 & culture & -0.17\\
& directed & 0.18 & son & 0.18 & close & -0.15 & htm & -0.16\\
& crime & 0.18 & boy & 0.18 & case & -0.15 & music & -0.15\\
& natural & 0.18 & biography & 0.17 & appointed & -0.15 & paris & -0.15\\

\hline
\end{tabular}
\caption{
Top positive and negative predictors (words) of reference clicks (\Secref{sec:refClickStudy}), for different article topics.
Words are organized based on where they appear: in the sentence annotated by the reference, or in the reference text.
}
\vspace{-15pt}
\label{tab:word_regression}
\end{table}

%% file: 6_RQ3.tex
\section{RQ3: Link-level analysis of citation interactions}
\label{sec:RQ3}

Our final research question asks which features of a specific reference predict if readers will engage with it.
Note that this is different from RQ2 (\Secref{sec:RQ2}), where we operated at the page level and did not differentiate between different references on the same page.

\subsection{Predictors of reference clicks}
\label{sec:refClickStudy}

We begin with a regression analysis to detect which
features predict whether a reference
will be clicked.
We selected all the references with external links, and we carefully rule out a host of confounds by sampling pairs of clicked and unclicked references from the same page view, thus controlling for situational features such as the page, user, information need, \etc{}
As we saw in \Figref{fig:click_pos}, references at the top and bottom of pages are \textit{a priori} more likely to be clicked.
Thus, to exclude position as a confound and maximize the probability that the user saw both references in a pair, we pick as the unclicked reference in a pair the one that appears closest in the page to the clicked reference.
To make sure we sample references associated with a sentence, we discard all footnotes in tables, infoboxes, and images, and keep only those within the article text.
Finally, we again sample only one pair per session in order to avoid over\hyp representing readers who are more prone to click on references. This process yields 1.8M reference pairs.

\begin{figure*}[t]
    \begin{minipage}[t]{.3\textwidth}
        \centering
        \includegraphics[width=\textwidth]{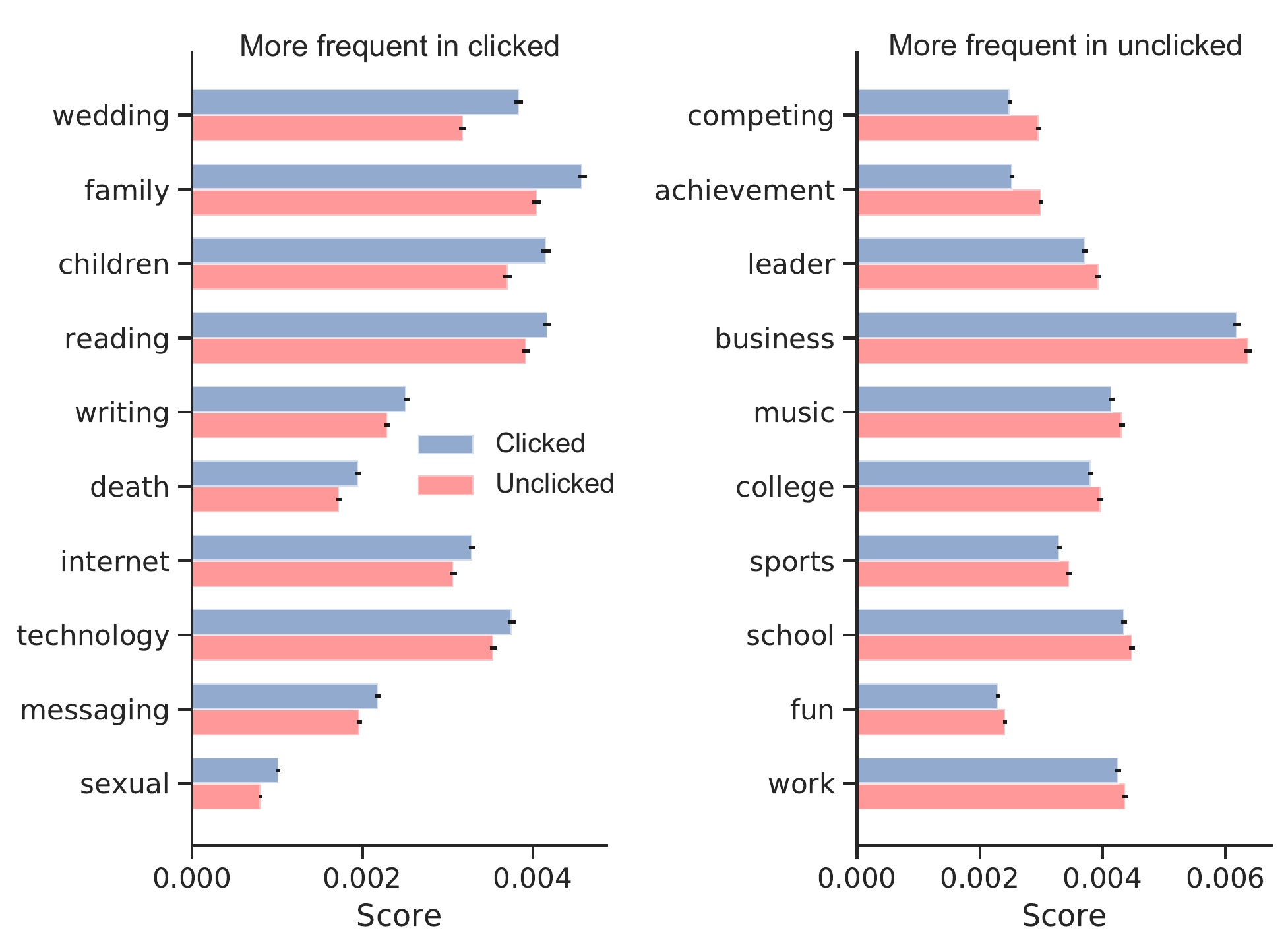}
        \subcaption{Click event (sentence text)}\label{fig:click_topics}
    \end{minipage}
    \hfill
    \begin{minipage}[t]{.3\textwidth}
        \centering
        \includegraphics[width=\textwidth]{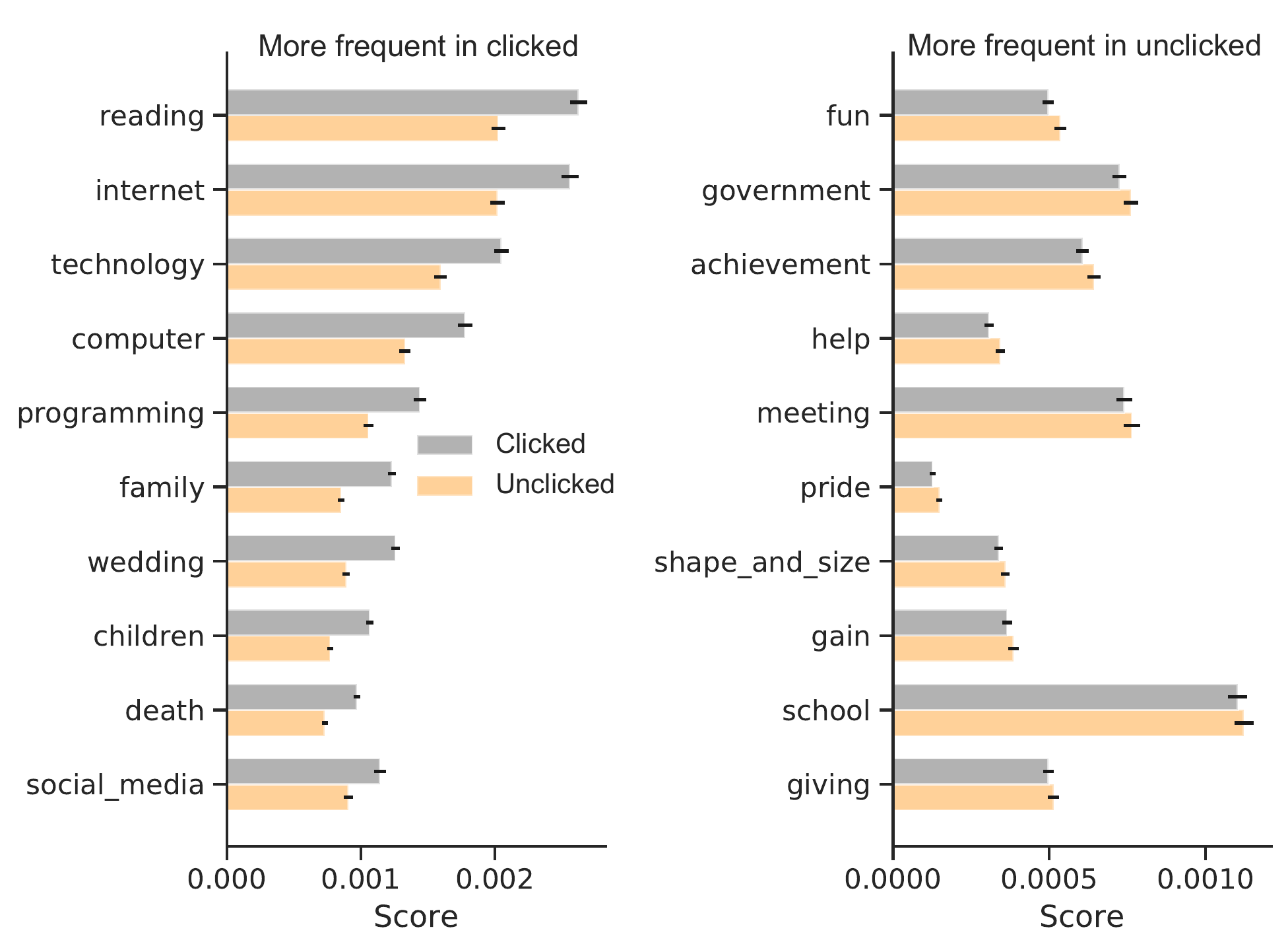}
        \subcaption{Click event (reference text)}\label{fig:click_topics_ref}
    \end{minipage}
    \hfill
    \begin{minipage}[t]{.3\textwidth}
        \centering
        \includegraphics[width=\textwidth]{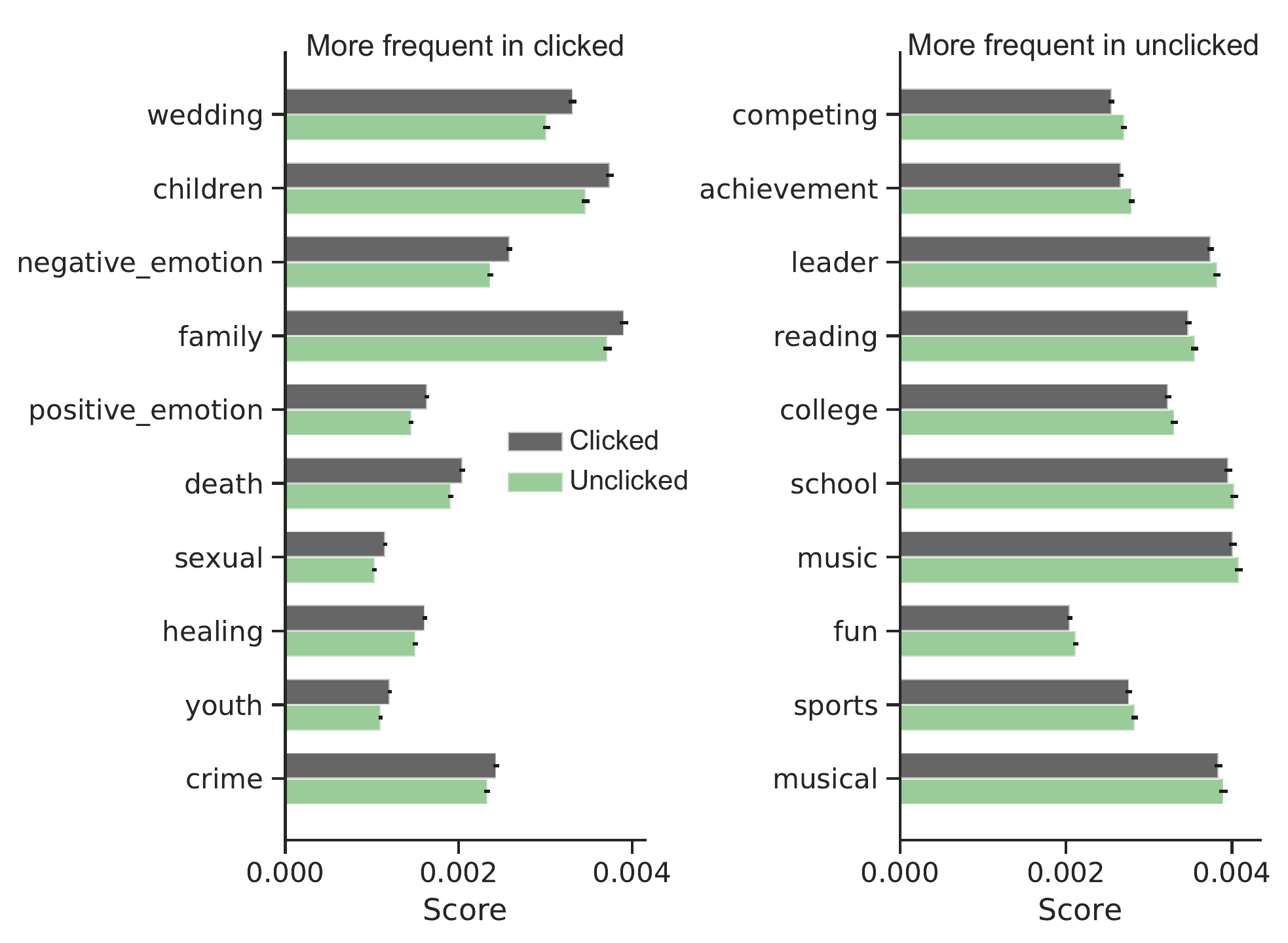}
        \subcaption{Hover event (sentence text)}\label{fig:hover_topics}
    \end{minipage}  
    \caption{\textit{Empath} \cite{Fast:2016:EUT:2858036.2858535} topics most strongly (anti-)associated with citation events (\cf\ \Secref{sec:refTopicClickStudy} for description).
    Reference text not studied for hover event (\Secref{sec:Predictors of footnote hovering}) because unlikely to be visible to user before hovering.}
    \label{fig:topics_distribution}
\end{figure*}

As predictors we use the words in the sentence that cites the respective reference, as well as the words in the reference text (\cf\ \Secref{sec:Background: Citations in Wikipedia}), represented as binary indicators specifying for each of the 1K most frequent words whether the word appears in the sentence.%
\footnote{Stop words were removed, and numbers (except for 4-digit numbers that potentially represent years) were converted to a special number token.}
Using these features as predictors, we train a logistic regression to predict the binary click indicator.

We perform this analysis on the full above\hyp described dataset, as well as on subsets consisting only of page views from each of 4 broad categories (derived by aggregating the 44 WikiProjects categories from \Secref{sec:Capturing event context}):
``Culture'' (1.3M pairs), ``STEM'' (436K), ``Geography'' (530K), and ``History and Society'' (467K).
The model achieves a testing AUC of around 0.55 across these 5 settings.

The words with the largest and smallest coefficients are displayed in \Tabref{tab:word_regression}, where we observe that, for all article topics except for ``STEM'', many positive features are related to social and life events and relationships (``dies'', ``obituary'', ``married'', ``wife'', ``relationship'', ``sex'', ``daughter'', ``family'', \etc).
Another common pattern across topics is that ``2019'' is strongly related with clicking,
and that career-related references (``awards'', ``debut'', \etc) are less likely to be clicked.
We shall further discuss these observations in \Secref{sec:Discussion}.

On STEM-related pages, open\hyp access references seem to receive more clicks than others, with words like ``free'' and ``pdf'' among the top predictors,
whereas words related to traditionally closed-access libraries such as JSTOR appear among the negative predictors, in line with previous findings~\cite{teplitskiy_amplifying_2017}.

\subsection{Topical correlates of reference clicks}
\label{sec:refTopicClickStudy}

For a higher-level view, we perform a topical analysis of citing sentences and reference texts, separately for the clicked \vs\ the unclicked references from the paired dataset of \Secref{sec:refClickStudy}.

To extract topics, we use \emph{Empath} \cite{Fast:2016:EUT:2858036.2858535}, which comes with a pre-trained model for labeling input text with a distribution over 200 wide-ranging topics.
After applying the model to each data point, we compute the average topic distribution for clicked and unclicked references, respectively,
and sort topics by the signed difference between their probability for clicked \vs\ unclicked references.

The topics with the largest positive and negative differences are listed in \Figref{fig:click_topics} and \ref{fig:click_topics_ref} for citing sentences and reference texts, respectively.
The results corroborate those from \Secref{sec:refClickStudy}, with human factors (wedding, family, sex, death) being more prominent among clicked references, whereas career-related topics such as competitions or achievements receive less attention.
Among the most prominent topics for reference texts (\Figref{fig:click_topics_ref}), topics related to technology and the Internet also emerge.

\subsection{Predictors of footnote hovering}
\label{sec:Predictors of footnote hovering}

The analyses of \Secref{sec:refClickStudy} and \ref{sec:refTopicClickStudy} considered engagement via reference clicks.
As we observed in \Figref{fig:event_distribution}, on desktop devices, hovering over a footnote to reveal the reference text in a tooltip is an even more common way to interact with references.
We hence replicated the above analyses with the \fnHover{} instead of the \refClick{} event (8.7M reference pairs), with the only difference that we excluded words from reference texts as features, since the user is unlikely to have seen those words before hovering over the footnote.

The results echo those of \Secref{sec:refClickStudy} and \ref{sec:refTopicClickStudy}, so for space reasons we do not discuss the regression analysis for footnote hovering (\cf\ \Secref{sec:refClickStudy}) and focus on the topical analysis instead (\cf\ \Secref{sec:refTopicClickStudy}).
Inspecting \Figref{fig:hover_topics}, we observe that we see a stronger tendency of \fnHover{} events, compared to \refClick{} events, to be elicited by words that are related to both positive and negative emotions.

\begin{table}
\footnotesize
\begin{tabular}{l r | l r} 

\multicolumn{2}{c|}{Positive}&
\multicolumn{2}{c}{Negative} \\

Word & Coeff. & Word & Coeff.\\
\hline
killer & 0.16 & oclc & -0.22\\
greatest & 0.16 & jason & -0.16\\
critic & 0.15 & episode & -0.15\\
things & 0.15 & die & -0.15\\
daughter & 0.15 & dictionary & -0.13\\
reveals & 0.14 & spanish & -0.12\\
baby & 0.14 & isbn & -0.12\\
instagram & 0.13 & le & -0.11\\
wife & 0.13 & board & -0.11\\
sheet & 0.13 & channel & -0.11\\

\end{tabular}\caption{
Top 10 positive and negative predictors (words)
of reference click following footnote hover (\Secref{sec:Predictors of reference clicks after hovering}).
}
\label{tab:sequence_table}
\vspace{-5mm}
\end{table}

\subsection{\hspace{-1mm}Predictors of reference clicks after hovering}
\label{sec:Predictors of reference clicks after hovering}

Once a user hovers over a  (\fnHover{}), the text of the corresponding reference is revealed in a so-called reference tooltip (\Figref{fig:event_types}).
At this point, the user has the choice to either click through to the citation URL (\refClick{}) or to stay on the article page.
In the final analysis of the paper, we are interested in understanding what words in the reference text influence the user when making this decision.

We create a dataset by selecting the page loads with at least two footnote hover events, where one converted to a \refClick{}
(positive), whereas the other did not (negative).
As in the previous studies, we selected at most one random pair per session, giving rise to a dataset of 440K pairs of hover events.

Similar to the study in \Secref{sec:refClickStudy}, we represent reference texts as 1K-dimensional word indicator vectors and use them as predictors in a logistic regression to predict \refClick{} events (testing AUC 0.54).

The strongest coefficients are summarized in \Tabref{tab:sequence_table}, painting a picture consistent with the previous analyses:
readers, after seeing a reference preview via the tooltip,
are more likely to click on the cited link when the reference text mentions social and life aspects (``wife'', ``baby'',  ``instagram'', \etc).
The strongest negative coefficients suggest that readers tend to not click through to dictionary entries, book catalogs (ISBN, OCLC), and information in languages other than English:
manual inspection revealed that ``spanish'' is mainly due to the note ``In Spanish'', ``le'' is the French article common in French newspaper names (\eg, \textit{Le Monde}),
and ``die'' is a German article.

%% file: 7_Discussion.tex
\section{Discussion and Conclusions}
\label{sec:Discussion}
Our analysis provides important insights regarding the role of Wikipedia as a gateway to information on the Web. 
We found that \textbf{in most cases Wikipedia is the final destination} of a reader's journey: fewer than 1 in 300 page views lead to a citation click. In our analysis, we focused on the fraction of users who engage with references, and characterized \textbf{how Wikipedia is used as a gateway to external knowledge.} Our findings suggest the following.

\begin{itemize}[leftmargin=*]
    \item
    \textbf{We engage with citations in Wikipedia when articles do not satisfy our information need.} \Secref{sec:RQ2} showed that readers are more likely to click citations on shorter and lower\hyp quality articles.
    Although this result seemed counter\hyp intuitive at first, since higher\hyp quality articles actually contain \textit{more} references that could potentially be clicked,
    it is in line with the finding that citations to sources reporting atomic facts that are typically available in Wikipedia articles (e.g., awards, career paths), are also generally less engaging (\Secref{sec:RQ3}).
    Collectively, these results suggest that readers are inclined to seek content beyond Wikipedia when the encyclopedia itself does not satisfy their information needs.
    \item
    \textbf{Citations on less engaging articles are more engaging.} In all of \Secref{sec:RQ2} we found that citation click-through rates decrease with the popularity of an article. While this may follow from the previous point because long, high\hyp quality articles tend to be more popular, it may also suggest that less popular articles are visited with a specific information need in mind. Previous work indeed suggests that popular articles are more likely to be viewed by users who are randomly exploring the encyclopedia \cite{singer_why_2017}.
    \item
    \textbf{We engage with content about people's lives.} We clearly saw that readers' interest is particularly high in references about people and their social and private lives (\Secref{sec:RQ3}).
    This is especially true for hovers, a less cognitively demanding form of engagement with citations. Hover events are also more likely to be elicited by words that are related to emotions, both positive and negative.
    \item
    \textbf{Recent content is more engaging.} We found that
    references about recent events (whose text includes ``2019'') are more engaging, both in terms of hovering and clicking.
    \item
    \textbf{Open content is more engaging.} Finally, we saw that references in Wikipedia pages about science and technology, especially if they point to a open\hyp access sources (e.g., having ``free'' or ``pdf'' in the reference text), are also more likely to be clicked.
\end{itemize}
\vspace{-8pt}

\xhdr{Theoretical implications}
Our findings furnish novel insights about Web users and their information needs through the lens of the largest online encyclopedia. For the first time, by characterizing Wikipedia citation engagement, we are able to quantify the value of Wikipedia as a gateway to the broader Web.
Our findings enable researchers to develop novel theories about readers' information needs and the possible barriers separating knowledge within and outside of the encyclopedia.
Our research can also guide the broader community of Web contributors in prioritizing efforts towards improving information reliability: we found that people especially rely on cited sources when seeking information about recent events and biographies, which suggests that Web content in these areas should be especially well curated and verified. Finally, the fact that readers engage more with freely accessible sources highlights the importance of open access and open science initiatives.

\xhdr{Practical implications}
Quantifying Wikipedia article completeness has proven to be a non-trivial task \cite{piscopo_what_2019}. The notion that article completeness is highly related to readers' engagement with Wikipedia references opens up ideas for novel applications to help satisfy Web users' information needs, including models that quantify lack of information in an article by incorporating signals related to reference click-through rate. Our findings will also help prioritize areas of content to be checked for citation quality by Wikipedia editors: in areas of content where Wikipedia acts as a major gateway, the quality and reliability of sources that readers visit become even more crucial. Finally, the data we collected could empower a model that, given a sentence missing a citation (i.e., with a \textit{citation needed} tag), could quantify how likely readers are to be interested in accessing the corresponding information and thereby help Wikipedia editors prioritize the backlog of unsolved missing\hyp reference cases.

\xhdr{Limitations and future work}
The overall low AUC (0.54 to 0.6) of the regression models (\Secref{sec:RQ2}--\ref{sec:RQ3})
emphasizes the inherent unpredictability of reader behavior.
While the significantly above-chance performance renders the models useful for analyzing the impact of various predictors, their performance is currently too low to make them useful as practical predictive tools.
Future work should hence invest in more powerful sequence models to improve accuracy.

By focusing on English Wikipedia only, the present analysis provides a limited view of the broader Wikipedia project, which is available in almost 300 languages and accessed by users all over the world. In our future work, we therefore plan to replicate this study for other language editions. 
So far, we also omitted any user characteristics from our study, such as more global behavioral traits beyond the page-view level, as well as geographic information, which are known to play an important role in user behavior
\cite{lemmerich_why_2019,teblunthuis_dwelling_2019}.
Future work should incorporate such signals.

We will also investigate reader intents more closely.
While click and hover logs
reflect the extent to which readers are interested in knowing more about a given topic,
they cannot tell us about the specific circumstances that led the user to engage by clicking or hovering,
nor about the level of satisfaction achieved by following up on a reference.
In the future, we plan to better understand these aspects via qualitative methods such as surveys and interviews.

Further, whereas our analysis focused on links in the \textit{References} section of articles, future work should also study other types of external links (\cf\ \Figref{fig:event_types}) in satisfying readers' information needs.

Finally, as exogenous events strongly affect Wikipedia users' information needs \cite{singer_why_2017}, future work should go beyond studying Wikipedia as an isolated platform and analyze how citation interaction patterns are warped by breaking news and events with uncertain information.
This will sharpen our picture of Wikipedia as a gateway to global information.